%% file: WWWPaper.tex
\newcommand*{\ATLASLATEXPATH}{latex/}
 
\newcommand\Tstrut{\rule{0pt}{2.6ex}}
 \documentclass[cernpreprint,texlive=2011,txfonts,UKenglish]{latex/atlasdoc} 
 \pdfoutput=1

\usepackage{\ATLASLATEXPATH atlaspackage}
\usepackage{\ATLASLATEXPATH atlasbiblatex}

\usepackage{\ATLASLATEXPATH atlascontribute}

\usepackage{\ATLASLATEXPATH atlasphysics}

\graphicspath{{logos/}{figures/}}

\usepackage{WWWPaper-defs}
\usepackage{placeins}
\usepackage{multirow}
\usepackage{pdflscape}
\usepackage{float}
\usepackage[utf8]{inputenc}  
\usepackage{biblatex}

\DeclareOldFontCommand{\rm}{\normalfont\rmfamily}{\mathrm}
\DeclareOldFontCommand{\sf}{\normalfont\sffamily}{\mathsf}
\DeclareOldFontCommand{\tt}{\normalfont\ttfamily}{\mathtt}
\DeclareOldFontCommand{\bf}{\normalfont\bfseries}{\mathbf}
\DeclareOldFontCommand{\it}{\normalfont\itshape}{\mathit}
\DeclareOldFontCommand{\sl}{\normalfont\slshape}{\@nomath\sl}
\DeclareOldFontCommand{\sc}{\normalfont\scshape}{\@nomath\sc}

\input{WWWPaper-metadata}

\hypersetup{pdftitle={ATLAS draft},pdfauthor={The ATLAS Collaboration}}

\begin{document}	

\maketitle
                                                                
\renewcommand{\figurename}{Figure}
\input{main.tex}

\section*{Acknowledgements}
\input{Acknowledgements.tex}

\printbibliography


\clearpage

\renewcommand{\figurename}{Figure}
\renewcommand{\tablename}{Table}

\newpage 
\input{atlas_authlist}

\end{document}

%% file: WWWPaper-metadata.tex
\AtlasTitle{Search for triboson \www\ production in $pp$ collisions at $\sqrt{s}=8$~\TeV\ with the ATLAS detector}

\author{The ATLAS Collaboration}

\AtlasRefCode{STDM-2015-07}
\PreprintIdNumber{CERN-EP-2016-172} 

\AtlasJournal{Eur. Phys. J. C}

\AtlasAbstract{%
This paper reports a search for triboson \www\ production in two decay channels \\(${\wwwlep}$ and ${\wwwsemilep}$ with $\ell=e, \mu$) in proton-proton collision data corresponding to an integrated luminosity of 20.3~$\ifb$ at a centre-of-mass energy of 8~\TeV{} with the ATLAS detector at the Large Hadron Collider. Events with exactly three charged leptons, or two leptons with the same electric charge in association with two jets, are selected. 
The total number of events observed in data is consistent with the Standard Model (SM) predictions. 
The observed 95\% confidence level upper limit on the SM \www\ production cross section is found to be 
730 fb with an expected limit of 560 fb in the absence of SM \www\ production.
Limits are also set on $WWWW$ anomalous quartic gauge couplings.
}

%% file: main.tex
\section{Introduction}
The triple gauge couplings (TGCs) and quartic gauge couplings (QGCs) that describe the strengths 
of the triple and quartic gauge boson self-interactions are completely determined by the non-Abelian nature 
of the electroweak SU(2)$_{\textrm{L}} \times $~U(1)$_{\textrm{Y}}$ gauge structure in the Standard Model (SM). 
These interactions contribute directly to diboson and triboson production at colliders. 
Studies of triboson production can test these interactions and any possible observed deviation from the theoretical prediction would provide hints of new physics at a higher energy scale.
Compared with TGCs, QGCs are usually harder to study due to the, in general, smaller production cross sections of the relevant processes. 

In the SM, charged QGC interactions ($WWWW$, $WWZZ$, $WWZ\gamma$ and $WW\gamma\gamma$) are allowed whereas neutral QGC interactions ($ZZZZ$, $ZZZ\gamma$, $ZZ\gamma\gamma$, 
$Z\gamma\gamma\gamma$ and $\gamma\gamma\gamma\gamma$) are forbidden. 
Searches have been performed by the LEP experiments for $WW\gamma\gamma$, $WWZ\gamma$, and $ZZ\gamma\gamma$ 
QGCs~\cite{delphi1, L3_1, L3_2, opal1, opal2, opal3}, by the Tevatron experiments for $WW\gamma\gamma$~\cite{d0}, 
and by the LHC experiments for $WW\gamma\gamma$, $WWZ\gamma$, $WWZZ$, $ZZ\gamma\gamma$, $Z\gamma\gamma\gamma$,  
and $WWWW$ QGCs~\cite{atlas1, atlas2,atlas3, atlas4, atlas5, atlas6, atlas7, cms1, cms2, cms4}.

Previous studies of $WWWW$ QGC interactions~\cite{atlas1, cms2} used $W^\pm W^\pm$ vector-boson scattering events, whereas this paper 
presents the first search for $WWWW$ QGC interactions via triboson \www\ production and sets the first limit on the total SM \www\ production cross-section using proton-proton ($pp$) collision data collected with the ATLAS detector and corresponding to an integrated luminosity of 20.3~$\ifb$~\cite{lumi2} at a centre-of-mass energy of 8~\TeV{}.
Two decay channels, \wwwlep\ and \wwwsemilep{}, with $\ell=e$ or $\mu$, are considered and are hereafter referred to simply as \3l\ and \2l2j channels, respectively. 

\section{The ATLAS detector}
The ATLAS detector~\cite{detector} is composed of an inner tracking detector (ID) surrounded by a thin
superconducting solenoid providing a 2~T axial magnetic field,
electromagnetic and hadronic calorimeters, and a muon spectrometer (MS). 
The ID consists of three subsystems: the pixel and silicon
microstrip detectors that cover $|\eta|<2.5$ in pseudorapidity\footnote{The ATLAS experiment uses a right-handed coordinate system with its origin at the nominal interaction point 
(IP) in the centre of the detector. The $x$-axis points from the IP to the centre of the LHC ring, the 
$y$-axis points upward, and the $z$-axis is along one of the proton beam directions. Cylindrical coordinates $(r,\phi)$ are used in the transverse plane, 
$\phi$ being the azimuthal angle around the beam pipe. The pseudorapidity is defined in terms of the polar 
angle $\theta$ as $\eta=-\ln\tan(\theta/2)$. Transverse momentum (\pt) 
is defined relative to the beam axis and is calculated as $\pt=p \sin \theta$ where $p$ is the momentum.}, and the outer transition radiation tracker that has an acceptance 
range of $|\eta|<2.0$. The finely-segmented electromagnetic calorimeter is composed of lead absorbers with liquid argon (LAr) as the active
material, spanning $|\eta|<3.2$. In the region $|\eta|<1.8$, a pre-sampler detector using a thin layer of LAr is used to
correct for the energy loss by electrons and photons upstream of the calorimeter.
The hadronic tile calorimeter ($|\eta|<1.7$) consists of steel absorbers and scintillating tiles and is located directly
outside the envelope of the barrel electromagnetic calorimeter.
The endcap hadronic calorimeters use LAr as active material, with copper as absorber material, 
while the forward calorimeters use LAr as active material, with copper absorber for the first layer, 
dedicated to electromagnetic measurements, and tungsten for other layers, dedicated to hadronic measurements.
The MS is composed of three large superconducting air-core toroidal magnets,
a system of three stations of tracking chambers in the range $|\eta|<2.7$,
and a muon trigger system in the range $|\eta|< 2.4$.
The precision muon momentum measurement is performed by monitored drift tubes 
everywhere except in the innermost layer for the range $|\eta| > 2.0$ where cathode strip chambers are used instead.
The muon trigger system is composed of resistive plate chambers in the barrel region ($|\eta| < 1.05$)
and thin gap chambers in the endcap region ($1.05 < |\eta| < 2.4$).

The ATLAS trigger system has three distinct levels referred to as L1, L2, and the event filter.
Each trigger level refines the decisions made at the previous level.
The L1 trigger is implemented in hardware and uses a subset of detector information to reduce the 
event rate to a design value of at most 75~kHz. The L2 and event filter are software-based trigger levels and together 
reduce the event rate to about 400~Hz. 

Events used were selected by single-lepton triggers with a transverse momentum, \pt{}, 
threshold of 24~\GeV{} for both muons and electrons, along with an isolation requirement. 
The single-lepton triggers are complemented with triggers having a higher \pt{} threshold 
(60~\GeV{} for electrons and 36~\GeV{} for muons) and no isolation requirement in order 
to increase the acceptance at high \pt. 

\section{Object reconstruction and event selection}
Each event is required to have at least one primary vertex reconstructed from at least three tracks with $\pt>400$~\MeV{}.
If there are multiple primary vertices reconstructed in the event due to additional $pp$ interactions (pile-up) 
in the same or a neighbouring bunch crossing, the vertex with the highest $\sum \pt^2$, calculated using all associated tracks, 
is taken as the primary collision vertex. The mean number of interactions per bunch crossing in this data set is $20.7$.

Electron candidates~\cite{electron1} are required to have $\pt>20$~\GeV{} and $|\eta| < 2.47$. Candidates within the transition region between 
the barrel and endcap calorimeters ($1.37 < |\eta| < 1.52$) are rejected. In addition, they must 
satisfy the {\it tight} quality definition described in Ref.~\cite{electron2012}. 
Muon candidates are reconstructed by combining tracks in the ID with tracks in the MS and have $\pt>20$~\GeV{} and $|\eta|<2.5$. 
The ID tracks associated with these muons must pass a number of quality requirements~\cite{muon}. 

To ensure that lepton candidates originate from the primary vertex, a requirement is placed on the
longitudinal impact parameter, $z_0$, multiplied by the sine of the track polar angle, $\theta$, 
such that the absolute value is smaller than 0.5~mm ($|z_0 \times \sin \theta|<0.5$~mm).  
A requirement is also placed on the transverse impact parameter, $d_0$, divided by its resolution ($\sigma_{d_0}$), such that $|d_0/\sigma_{d_0}|<3$.
To suppress the contribution from hadronic jets which are misidentified as leptons, signal leptons
are required to be isolated in both the ID and the calorimeter.  The calorimeter isolation is defined as $\et^{\mathrm{Cone}X}/\et$ whereas 
the ID isolation is defined as $\pt^{\mathrm{Cone}X}/\pt$, where $\et^{\mathrm{Cone}X}$ ($\pt^{\mathrm{Cone}X}$) is the transverse energy (momentum)
deposited in the calorimeter (the scalar sum of the \pt\ of tracks with $\pt>1$~\GeV{}) within a cone of size $\Delta R=\sqrt{(\Delta \eta)^2 + (\Delta \phi)^2}=X$ around the lepton. 
The transverse momentum from the lepton itself is excluded in the calculations of $\et^{\mathrm{Cone}X}$ and $\pt^{\mathrm{Cone}X}$.
Different lepton isolation criteria are applied in the two channels to maximize the signal efficiency while suppressing the backgrounds. 
In the \3l\ channel, $\et^{\mathrm{Cone}0.2}/\et<0.1$ and $\pt^{\mathrm{Cone}0.2}/\pt<0.04$ are required for both the electrons and muons; 
in the \2l2j\ channel, $\et^{\mathrm{Cone}0.3}/\et<0.14$ and  $\pt^{\mathrm{Cone}0.3}/\pt<0.06$ are required for electrons whereas
$\et^{\mathrm{Cone}0.3}/\et<0.07$ and $\pt^{\mathrm{Cone}0.3}/\pt<0.07$ are required for muons.

Jets are reconstructed from clusters of energy in the calorimeter using the anti-$k_t$ algorithm~\cite{anti-kT} with radius parameter $R = 0.4$. 
Jet energies are calibrated using energy- and $\eta$-dependent correction factors derived using Monte Carlo (MC) 
simulation and validated by studies of collision data~\cite{jetenergy}. For jets with $\pt<50$~\GeV{} and $|\eta|<2.4$, at least 50\% of the summed scalar \pt~of the tracks 
within a cone of size $\Delta R=0.4$ around the jet axis must originate from the primary vertex.
This requirement reduces the number of jet candidates originating from pile-up vertices.
Jets containing $b$-hadrons (``$b$-jets'') with $|\eta| < 2.5$ and $\pt>25~\GeV$ are identified 
using the impact parameter significance of tracks in the jet and secondary vertices reconstructed from these tracks~\cite{btag1,btag2}. 
In the \3l\ and \2l2j\ channels, the efficiency of the $b$-tagging algorithm used is 85\% and 70\%, respectively. 

The measurement of the two-dimensional missing transverse momentum vector, \vecmet, is based 
on the measurement of all topological clusters in the calorimeter and muon tracks reconstructed in the ID and MS~\cite{met}. 
Calorimeter cells associated with reconstructed objects, such as electrons, photons, hadronically decaying $\tau$ leptons, and jets, are calibrated at their own energy scale, 
whereas calorimeter cells not associated with any object are calibrated at the electromagnetic energy scale and taken into account 
as a so-called ``soft term'' in the calculation of \vecmet.
The magnitude of the missing transverse momentum vector is referred to as the missing transverse energy, $\met = |\vecmet|$.

The experimental signature of the \3l\ channel is the presence of three charged leptons and \met{}. The signature of 
the \2l2j\ channel is the presence of two same-charge leptons, \met, and two jets with an invariant mass close to 80~\GeV{}. 
The selection requirements used to define the signal regions described in the following 
are obtained from a multi-dimensional optimization to maximize the sensitivity to the \www\ process 
and to reduce the contributions from SM background processes.

To select \3l\ candidates, events are required to have exactly three charged leptons with $\pt>20$~\GeV{}, 
at most one jet with $\pt>25$~\GeV{} and $|\eta|<4.5$, and no identified $b$-jets. 
In addition, the absolute value of the azimuthal angle between the trilepton system and the \vecmet, $|\phi^{3\ell} - \phi^{\vecmet}|$, is required to be above 2.5.
Eight different final states with equal production probability are considered based on the flavour and the charge of the leptons, namely $e^\pm e^\pm e^\mp$,
$e^{\pm}e^{\mp}\mu^{\pm}$, $e^{\pm}e^{\mp}\mu^{\mp}$, $e^\pm e^\pm \mu^\mp$, $\mu^{\pm}\mu^{\mp}e^{\pm}$,
$\mu^{\pm}\mu^{\mp}e^{\mp}$, $\mu^\pm \mu^\pm e^\mp$, and $\mu^\pm \mu^\pm \mu^\mp$.
Three separate signal regions are defined based on the number of same-flavour opposite-sign (SFOS) lepton pairs in the event: 
0~SFOS ($e^\pm e^\pm \mu^\mp$ and $\mu^\pm \mu^\pm e^\mp$), 
1~SFOS ($e^{\pm}e^{\mp}\mu^{\pm}$, $e^{\pm}e^{\mp}\mu^{\mp}$, $\mu^{\pm}\mu^{\mp}e^{\pm}$, and $\mu^{\pm}\mu^{\mp}e^{\mp}$), 
and 2~SFOS ($e^\pm e^\pm e^\mp$ and $\mu^\pm \mu^\pm \mu^\mp$). 
In the 0-SFOS case, the invariant mass of the same-flavour lepton pair, $m_{\ell\ell}$, is required to be greater than 20~\GeV{}. If there are at least two electrons in the event, 
the di-electron invariant mass, $m_{ee}$, is required to have $|m_{ee} - m_Z|>15$~\GeV{}, where 
$m_Z$ is the pole mass of the $Z$ boson~\cite{pdg}. No requirement is applied on the $\met$ variable, 
as it was found to not discriminate between signal and backgrounds.  
In the 1-SFOS case, the SFOS dilepton invariant mass, $m_{\rm SFOS}$, is required to be outside of the region 
$m_Z-35~\GeV{}<m_{\text{SFOS}}<m_Z+20$~\GeV{}. 
In addition, events are required to satisfy $\met>45$~\GeV{}.
Finally, in the 2-SFOS case, the SFOS dilepton invariant masses are required to have $|m_{\rm SFOS}-m_Z|>20$~\GeV{} 
while the \met~must be greater than 55~\GeV. The selection criteria for $m_{\rm SFOS}$ and \met\ 
are mainly used to reduce the contributions from the $Z+$jets and $WZ+$jets processes. Table~\ref{tab:cut_3l} shows the kinematic selection criteria used for the \3l\ channel.

\begin{table*}
\begin{center}
	\caption{Selection criteria for the \3l\ channel, split based on the number
	of SFOS lepton pairs: 0 SFOS, 1 SFOS, and 2 SFOS.} 
	\label{tab:cut_3l}
	\small
\begin{tabular}{l|c|c|c} 
	  \hline
\3l\ & \multicolumn{1}{c|}{0~SFOS} & \multicolumn{1}{c|}{1~SFOS}    & \multicolumn{1}{c}{2~SFOS}  \\ \hline\hline
Preselection & \multicolumn{3}{c}{Exactly three charged leptons with $\pt>20$~\GeV{}} \\ \hline
\met &  \multicolumn{1}{c|}{-} & \multicolumn{1}{c|}{$\met>45$~\GeV{}} & \multicolumn{1}{c}{$\met>55$~\GeV{}}  \\  \hline
Same-flavour dilepton mass        & \multicolumn{1}{c|}{$m_{\ell\ell}>20$~\GeV{}} &  \multicolumn{2}{c}{-} \\ \hline
Angle between trilepton and \vecmet        & \multicolumn{3}{c}{$|\phi^{3\ell} - \phi^{\vecmet}|>2.5$} \\ \hline
\multirow{3}{*}{$Z$ boson veto} &  \multirow{3}{*}{$|m_{ee}-m_Z|>15$~\GeV{}} & $m_Z - m_{\text{SFOS}} > 35$~\GeV{} & \multirow{3}{*}{$|m_{\text{SFOS}}-m_Z|>20$~\GeV{}} \\
&  & or & \\
&  & $m_{\text{SFOS}} - m_Z > 20$~\GeV{} & \\ \hline
Jet veto                                    & \multicolumn{3}{c}{At most one jet with $\pt>25$~\GeV{} and $|\eta|<4.5$} \\ \hline
$b$-jet veto                              & \multicolumn{3}{c}{No identified $b-$jets with $\pt>25$~\GeV{} and $|\eta|<2.5$} \\
\hline
\end{tabular}
\end{center}
\end{table*}

To select \2l2j\ candidates, events are required to have exactly two leptons with the same electric charge, at least two jets, and no identified $b$-jets. 
Three different final states are considered based on the lepton flavour, namely \ssee, \ssem, and \ssmm.
The lepton \pt~(\met) threshold is set to 30 (55)~\GeV{} to reduce the SM background contributions, though
the \met\ criterion is not applied for the \ssmm\ final state due to the smaller $Z+$jets background expected in this channel. 
The leading (sub-leading) \pt\ jet must have $\pt>30$ (20)~\GeV{} and $|\eta|<2.5$. The two jets are required to have $65~\GeV{}<m_{jj}<105$~\GeV{} and $|\Delta \eta_{jj}|<1.5$ 
in order to distinguish the signal from the $W^\pm W^\pm$ backgrounds, 
where $m_{jj}$ is the dijet invariant mass and $\Delta \eta_{jj}$ is the pseudorapidity separation between the two jets.   
The dilepton system is required to have $m_{\ell\ell}>40$~\GeV{} and in the case of the \ssee\ final state, 
$m_{ee}$ must have $m_{ee}<80$~\GeV\ or $m_{ee}>100$~\GeV\ in order to suppress events with two opposite-sign prompt leptons where the charge of one of the electrons is misidentified.
To reduce the contributions from $WZ+$jets and $ZZ+$jets production, events are
removed if they contain additional leptons reconstructed with $\pt>6$~\GeV{} passing looser identification quality requirements, 
with a {\it medium} identification requirement for electrons as defined in Ref.~\cite{electron2012}
and the minimum identification required for muon reconstruction.
Table~\ref{tab:cut_2l2j} shows the kinematic selection criteria used for the \2l2j\ channel.

\begin{table*}
\caption{Selection criteria for the \2l2j\ channel, split based on the lepton flavour: \ssee{}, \ssem{}, and \ssmm{}.}	
\begin{center}
\begin{tabular}{l|ccc}   \hline
\2l2j\ & \multicolumn{1}{c|}{\ssee} & \multicolumn{1}{c|}{\ssem}    & \multicolumn{1}{c}{\ssmm}  \\ \hline\hline
Lepton & \multicolumn{3}{c}{Exactly two same-charge leptons with $\pt>30$~\GeV{}} \\ \hline
Jets        & \multicolumn{3}{c}{At least two jets with $\pt(1)>30$~\GeV{}, $\pt(2)>20$~\GeV{} and $|\eta|<2.5$} \\ \hline
$m_{\ell\ell}$        & \multicolumn{3}{c}{$m_{\ell\ell}>40$~\GeV{}}  \\ \hline
\met                      & \multicolumn{2}{c|}{~~~~~~~~~~~~~~~~~~~~~$\met>55$~\GeV{}~~~~~~~~~~~~~~~~~~~~~~~~} & \multicolumn{1}{c}{-}  \\  \hline
$m_{jj}$        & \multicolumn{3}{c}{$65~\GeV<m_{jj}<105$~\GeV{}}  \\ \hline
$\Delta \eta_{jj}$        & \multicolumn{3}{c}{$|\Delta \eta_{jj}|<1.5$}  \\ \hline
\multirow{2}{*}{$Z$ boson veto}        & \multicolumn{1}{c|}{$m_{ee}<80$~\GeV{} or}  & \multicolumn{2}{c}{\multirow{2}{*}{-}}  \\ 
                                                   & \multicolumn{1}{c|}{$m_{ee}>100$~\GeV{}}    \\ \hline
Third-lepton veto                      & \multicolumn{3}{c}{No third lepton with $\pt>6$~\GeV{} and $|\eta|<2.5$ passing looser identification} \\ 
                      & \multicolumn{3}{c}{requirements}\\\hline
$b$-jet veto                              & \multicolumn{3}{c}{No identified $b-$jets with $\pt>25$~\GeV{} and $|\eta|<2.5$} \\
\hline
\end{tabular}
\end{center}
\label{tab:cut_2l2j}
\end{table*}

\section{Signal fiducial cross sections}
\label{sec:signal}

At leading order (LO), the production of three $W$ bosons can take place through radiation from a fermion, 
from an associated $W$ and $Z/\gamma^*/H$ production with the intermediate $Z/\gamma^*/H$ boson decaying to two opposite-sign $W$ bosons, 
or from a $WWWW$ QGC vertex. Representative Feynman graphs for each of these production processes are 
shown in Figure~\ref{fig:Sig-Fey-diagrams}. 
Calculations are available including corrections at next-to-leading order (NLO) in QCD
with all spin correlations involved in the vector-boson decays,
the effects due to intermediate Higgs boson exchange, and off-shell contributions correctly taken 
into account~\cite{Campanario:2008yg}. 
Electroweak NLO corrections have been calculated recently~\cite{Yong-Bai:2016sal}. However, they are 
not considered in this analysis.

In order to determine \www\ production cross sections, events are 
generated at NLO in QCD using \MGMCatNLO~\cite{madgraph_amc} including on-shell diagrams as well as Higgs associated diagrams. 
The CT10 NLO parton distribution function (PDF)~\cite{pdf} is used. 
Subsequent decays of unstable particles and parton showers are handled by {\sc pythia8}~\cite{pythia}. Fiducial cross sections are calculated using
the generator-level lepton, jet, and \met{} definitions as described in Ref.~\cite{ATL-PHYS-PUB-2015-013}.
Generator-level prompt leptons (those not originating from hadron and $\tau$ lepton decays) 
are dressed with prompt photons within a cone of size $\Delta R = 0.1$. 
Generator-level jets are reconstructed by applying the anti-$k_t$ algorithm with 
radius parameter $R=0.4$ on all final-state particles after parton showering and hadronisation. 
The \met\ variable is calculated using all generator-level neutrinos.
The same kinematic selection criteria as listed in Table~\ref{tab:cut_3l} and Table~\ref{tab:cut_2l2j} are applied on these objects, with the exception of the $b$-jet veto requirements in the $\ell \nu \ell \nu \ell \nu$ channel and the lepton quality requirements. To take into account the effect of the lepton isolation in the fiducial region, any lepton pairs must satisfy $\Delta R(\ell, \ell)>0.1$, and in the $\ell \nu \ell \nu jj$ channel any lepton-jet pairs must satisfy $\Delta R(j, \ell)>0.3$.
Electrons or muons from $\tau$ decays are not included. 

\begin{figure}[h!]
\begin{center}
\includegraphics[width=0.3\textwidth]{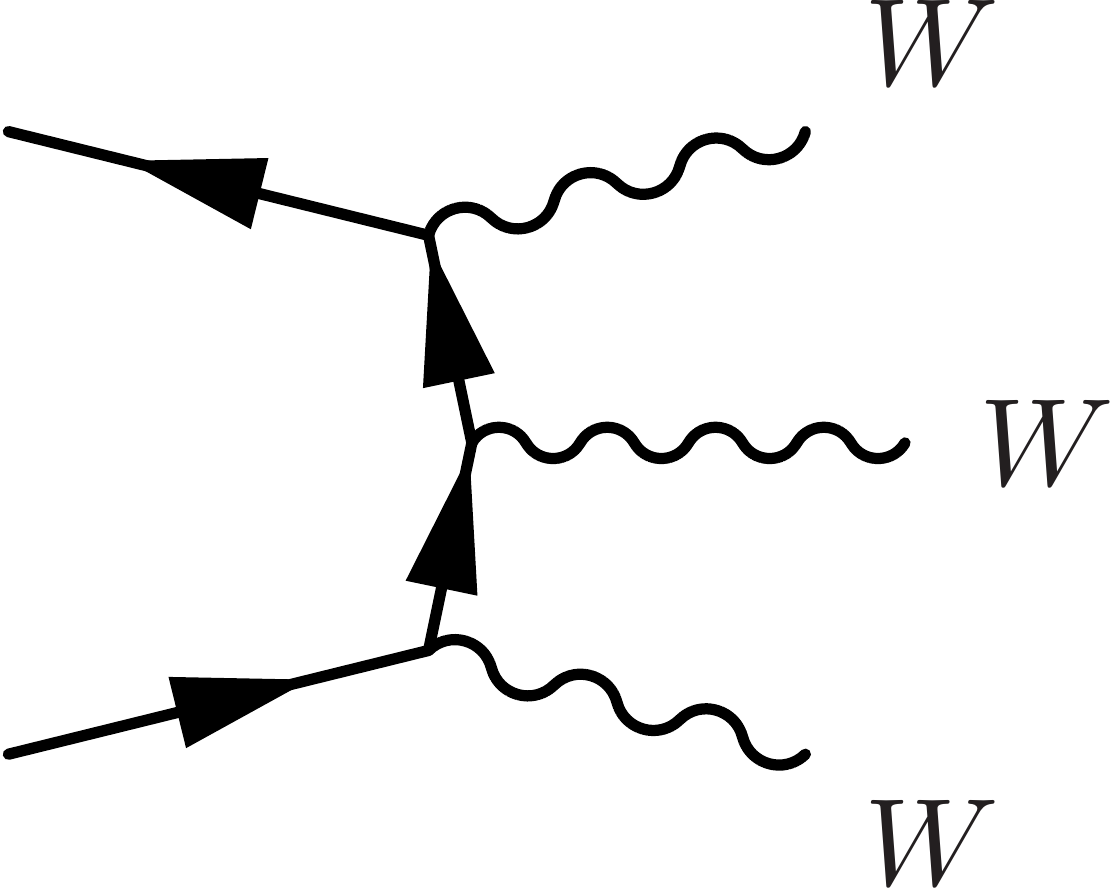}
\includegraphics[width=0.3\textwidth]{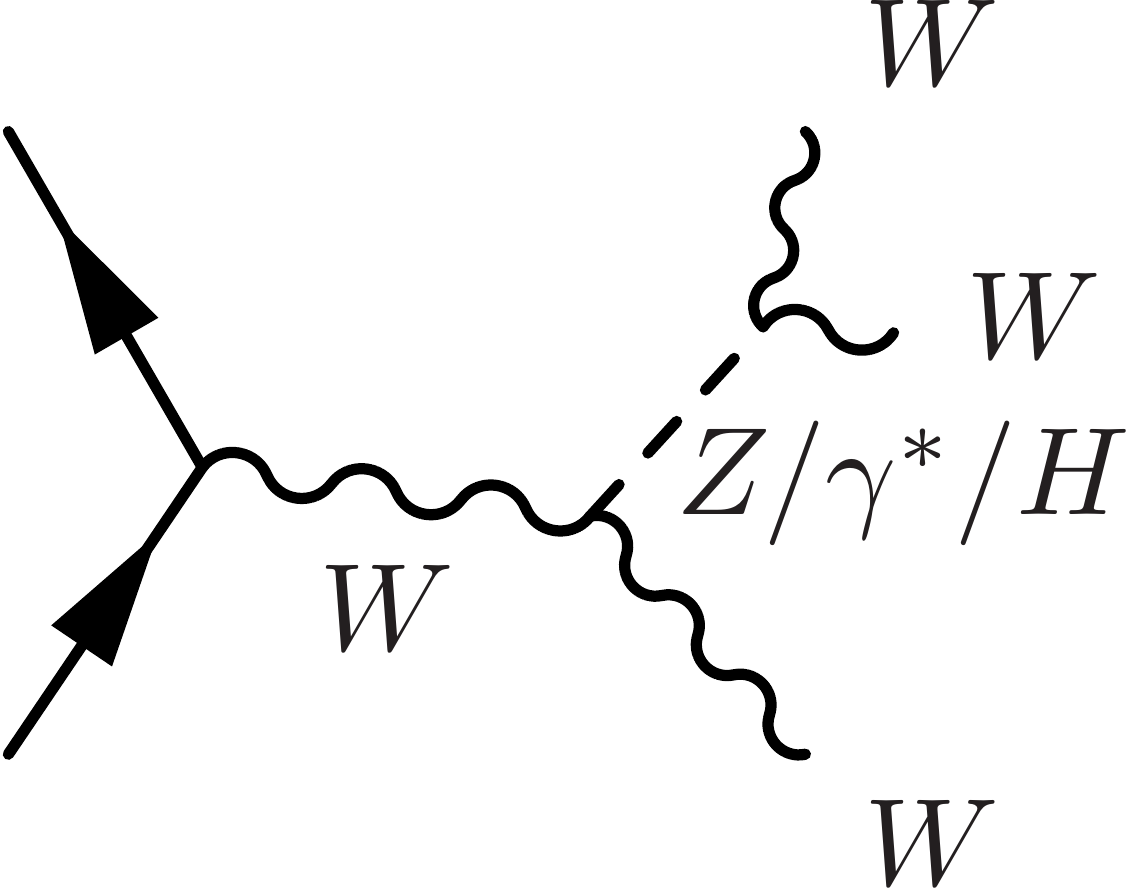} 
\includegraphics[width=0.3\textwidth]{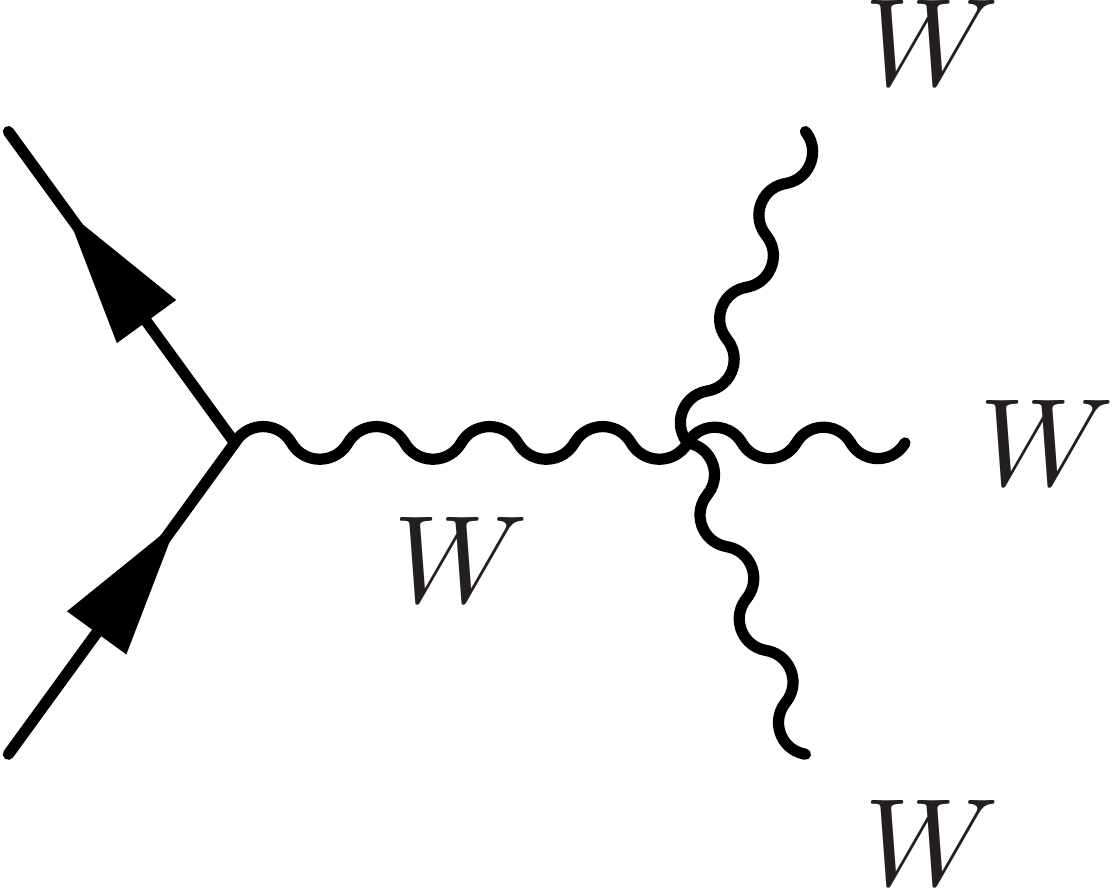} 
\end{center}
\caption{Feynman graphs contributing at LO to \www\ production.}
\label{fig:Sig-Fey-diagrams} 
\end{figure}

The fiducial cross section is predicted to be 
\mbox{$309 \pm 7~({\rm stat.}) \pm{}15~({\rm PDF}) \pm 8~({\rm scale})$} ab 
in the \3l\ channel and 
\mbox{$286 \pm 6~({\rm stat.}) \pm{}15~({\rm PDF}) \pm 10~({\rm scale})$} ab
in the \2l2j\ channel.
Uncertainties due to the PDFs are computed using an envelope of the {\sc CT10}, {\sc NNPDF3.0}~\cite{nnpdf}, 
and {\sc MSTW2008}~\cite{mstw} NLO PDF 68\% or 90\% (for {\sc CT10}) confidence level (CL) uncertainties, 
following the recommendation of Ref.~\cite{Alekhin:2011sk}.
The renormalization and factorization scales are set to the invariant mass of the $WWW$ system.
Scale uncertainties are estimated by varying the two scales independently up and down by a factor 
of two and taking the largest variation from the nominal cross-section values.

In order to combine the measurements from the two decay channels, 
a common phase space is defined 
where each $W$ boson can 
decay either leptonically (including $\tau$ leptons) or hadronically, $pp\rightarrow W^\pm W^\pm W^\mp$ $+~X$, 
with no kinematic requirements placed on the final-state leptons
but with jets restricted to have $\pt > 10$~\GeV{}. The extrapolation factor from the fiducial phase space 
to the total phase space is large, but it is mainly due to the well-known $W$ boson decay branching ratios. 
The total cross section in this common phase space is 
 $241.5 \pm 0.1$~(stat.)~$\pm 10.3$~(PDF) $\pm 6.3$~(scale)~fb.

In order to determine the detector reconstruction effects on the signal selection,
\www\ signal samples are generated with {\sc vbfnlo}~\cite{Campanario:2008yg, vbfnlo,vbfnlo2,vbfnlo3} at LO. The 
parton shower and hadronisation are performed by {\sc pythia8}. The fiducial cross sections are
seen to be consistent between {\sc vbfnlo} and \MGMCatNLO\ when computed at the same order. 
The {\sc vbfnlo} LO fiducial cross sections are 
normalized to the NLO fiducial cross section predicted by \MGMCatNLO\ for the signal yield calculations.
These events are processed through the full ATLAS detector simulation~\cite{det-sim} based on \GEANT4~\cite{geant}. 
To simulate the effect of multiple $pp$ interactions occurring during the same or a neighbouring bunch crossing, 
minimum-bias interactions are generated and overlaid on the hard-scattering process. 
These events are then processed through the same object reconstruction and identification algorithms as used on data. 
MC events are reweighted so that the pile-up conditions in the simulation match the data.
Additional corrections are made to the simulated samples to account for small differences between the simulation
and the data for the object identification and reconstruction efficiencies, the trigger efficiencies, 
and the energy and momentum scales and resolutions.
While excluded in the fiducial cross-section definition, the contribution from events
with $W\rightarrow \tau \nu\rightarrow\ell\nu\nu\nu$ decays are counted as signal in the {\sc vbfnlo}
signal sample used in the final event selection. These events contribute up to 20\% of the predicted signal yield. This approach is used to ease comparisons of the obtained cross-section limits with alternative cross-section predictions that may not simulate tau decays.

\section{Backgrounds}
\label{sec:bg}
\subsection{Background estimation}
\label{sec:bg_est}
The SM processes that mimic the \www\  signal signature can be grouped into five categories:
\begin{itemize}
\item The $WZ/\gamma^*+$jets process 
that produces three prompt leptons or two prompt leptons with the same electric charge (referred to as ``$WZ$ background''); 
\item The $W\gamma+$jets or $Z\gamma+$jets processes where the photon is misreconstructed as a lepton (referred to as ``$V\gamma$ background'', where $V=W, Z$); 
\item Processes other than $WZ/\gamma^*+$jets that produce three prompt leptons or two prompt leptons with the same electric charge (referred to as ``other prompt background''); 
\item Processes that produce two or three prompt charged leptons, but the charge of one lepton is misidentified (referred to as ``charge-flip background''); 
\item Processes that have one or two non-prompt leptons originating either from misidentified jets or from hadronic decays (referred to as ``fake-lepton background'').
\end{itemize}

The dominant irreducible background originates from the $WZ(\rightarrow \ell^\pm \nu\ell^\pm \ell^\mp)+$jets process 
and is estimated using simulated events. In the \3l\ channel these events are generated with \POWHEG{}-{\sc BOX}~\cite{Nason:2004rx,Frixione:2007vw,Alioli:2010xd,Melia:2011tj} 
and hadronised with {\sc pythia8} and in the \2l2j\ channel they are generated with \SHERPA{}~\cite{sherpa}. 
In the \3l\ channel, the inclusive $WZ+$jets cross section is normalized using a scale factor ($1.08\pm0.10$) derived 
from a $WZ$-enriched region in data. This region is obtained by requiring exactly one SFOS lepton pair with 
$|m_{\rm SFOS}-m_Z|<15~\GeV$. In the \2l2j\ channel, the cross section is normalized to 
the NLO calculation in QCD from {\sc vbfnlo}~\cite{Campanario:2010hp} in the specified fiducial phase space with a normalization factor of $1.04\pm0.09$.

The $V\gamma$ background contributes when the photon is misidentified as an electron. In the \3l\ channel,
this originates primarily from the $Z\gamma$ process and its contribution is estimated using events generated with \SHERPA{}. 
In the \2l2j\ channel, this comes primarily from electroweak and strong production of $W\gamma jj$ events.  
Strong production of $W\gamma jj$~\cite{Campanario:2014dpa} is estimated using \ALPGEN{}~\cite{alpgen} 
interfaced to \HERWIG{}~\cite{herwig} and \JIMMY{}~\cite{jimmy} for simulation of the parton shower, fragmentation, hadronisation and the underlying event. 
The electroweak production of $W\gamma jj$~\cite{Campanario:2013eta} is modelled using \SHERPA{}.

Other SM processes that produce multiple prompt leptons include $ZZ$, $t\bar{t}V$, $ZWW$, $ZZZ$, $W^\pm W^\pm jj$ production,
and double parton scattering processes. 
The production of $ZZ$ is modelled with \POWHEG{}-{\sc BOX}~\cite{Melia:2011tj} and hadronised with {\sc pythia8} 
in the \3l\ channel and is modelled with \SHERPA{} in the \2l2j\ channel.
The $t\bar{t}V$~\cite{Frixione:2015zaa}, $ZWW$~\cite{Binoth:2008kt}, and $ZZZ$~\cite{Binoth:2008kt}
processes are modelled using \MGMCatNLO\ together with {\sc pythia8} for both channels.
The non-resonant $W^\pm W^\pm jj$ background~\cite{Campanario:2013gea} is only important for the \2l2j\ channel
and its contribution is estimated using \SHERPA.
Contributions from double parton scattering processes are found to be negligible in both channels. 

The charge-flip background originates from processes where the charge of at least one prompt lepton is misidentified.
This occurs primarily when a lepton from a hard bremsstrahlung photon conversion is recorded instead of the signal lepton.
It mainly contributes to the 0-SFOS signal region in the \3l\ channel and the \ssee\ / \ssem\ signal regions in the \2l2j\ channel. 
The electron charge misidentification rate is measured using $Z \rightarrow e^+e^-$ events.  
In the \3l\ channel, the charge-flip background is estimated by using these rates to re-weight the MC estimate of $WZ$ and 
$ZZ$ events based on the probability for opposite-sign events of this kind to migrate into the 0-SFOS category. 
In the \2l2j\ channel, the background is estimated by applying these rates on data events satisfying all signal selection 
criteria except the two leptons are required to have opposite-sign.

Contributions from fake-lepton backgrounds are estimated in data, using different approaches in the two channels. 
In the \3l\ channel, the probabilities of prompt leptons or non-prompt leptons to satisfy the signal lepton criteria are computed using a tag-and-probe method 
whereby a well-reconstructed ``tag'' lepton is used to identify the event and a second ``probe'' lepton is used to study the probabilities without bias. 
A tag lepton must satisfy the signal lepton requirements while a looser lepton selection criterion is defined for probe leptons with the lepton isolation 
requirements removed and the electron quality requirement loosened to {\it medium} as defined in Ref.~\cite{electron2012}. The probability for a prompt 
lepton to satisfy the signal lepton criteria is estimated using $Z \rightarrow \ell^+ \ell^-$ events with 
the tag-and-probe lepton pair required to have the same-flavour, opposite-sign and an invariant mass within 10 \GeV\ of the pole mass of the $Z$ boson. The probability for 
a non-prompt lepton from hadronic activity to satisfy the signal lepton requirement is estimated using the tag-and-probe method 
in a $W+$jets-enriched region with $\met > 10$ \GeV, the tag lepton is a muon with 
$\pt > 40$ \GeV, and the tag and probe leptons have the same electric charge. 
The probabilities are calculated separately for electrons and muons.  
A loosely identified set of data is also selected by requiring at least three loose leptons as defined above. This set of data, 
along with these probabilities are then used to estimate the background in the signal region with the matrix method~\cite{matrixMethod}.  

In the \2l2j\ channel, events that contain one signal lepton and one ``lepton-like'' jet are selected. A ``lepton-like'' jet satisfies all signal lepton 
selection criteria except that the isolation requirements are $0.14<\et^{\mathrm{Cone}0.3}/\pt<2$ and 
$0.06<\pt^{\mathrm{Cone}0.3}/\pt<2$ for electrons, 
and $0.07<\et^{\mathrm{Cone}0.3}/\pt<2$ and $0.07<\pt^{\mathrm{Cone}0.3}/\pt<2$ for muons. 
In addition, the $|d_0/\sigma_{d_0}|$ and $|z_0 \times \sin \theta|$ 
selection criteria are loosened to 10~mm and 5~mm, respectively. 
These events are dominated by non-prompt leptons and are scaled by a fake factor to estimate the non-prompt background.
The fake factor is the ratio of the number of jets satisfying the signal lepton identification criteria to the number of jets satisfying the ``lepton-like'' jet criteria. It
is measured as a function of the jet \pt\ and $\eta$ from a dijet-enriched sample 
selected by requiring a lepton back-to-back with a jet ($\Delta \phi_{j\ell} > 2.8$)  and $\met < 40$~\GeV{}.

\subsection{Validation of background estimates}
The background predictions are tested in several validation regions (VRs). 
These VRs are defined to be close to the signal region with a few selection criteria removed or inverted. They generally 
have dominant contributions from one or two background sources and a negligible contribution from the signal process.  
The signal and background predictions are compared to data for each VR in Table~\ref{tab:CR}.

In the \3l\ channel, three VRs are considered. 
The first VR, called the pre-selection region, tests the modelling of the $WZ$+jets background by requiring exactly three signal leptons. 
The distribution of the trilepton transverse mass,
$m_{T}^{3\ell} = \sqrt{2 \pt^{3\ell}\MET\Big(1-\cos(\phi^{3\ell} - \phi^{\vecmet})\Big)}$ where $\pt^{3\ell}$ is the \pt\  of the trilepton system, is shown at 
the top left of Figure~\ref{fig:Val}. This VR includes the three signal regions (0, 1, and 2~SFOS), 
but the effect of the signal is considered negligible at this stage of the selection, as shown in Table~\ref{tab:CR}. 
The $WZ$+jets purity is estimated to be around $70\%$ in this region.
The second region, called the fake-lepton region, tests the modelling of the fake-lepton background by requiring exactly three signal leptons 
with no SFOS lepton pairs and at least one $b-$jet. 
The distribution of the jet multiplicity, $N_{\textrm{jet}}$, is shown at the top right of Figure~\ref{fig:Val}. 
The purity of the fake-lepton background is estimated to be around $80\%$ in this region.
The third region, called $Z\gamma$ region, tests the modelling of the $Z\gamma$ background by requiring the presence of 
only $\mu^+\mu^- e^\pm$ events where the trilepton invariant mass is close to the $Z$ resonance peak. This restricts the 
main contributions to originate from the $Z\gamma \rightarrow \mu^+ \mu^- \gamma$ 
and $Z \rightarrow \mu^+ \mu^- \rightarrow \mu^+ \mu^- \gamma$ processes. 
The $Z\gamma$ purity is estimated to be around $70\%$ in this region.
The data are seen to be well described by the background in all three VRs.

In the \2l2j\ channel, five VRs are considered. 
The modelling of the charge-flip background is tested using \ssee\ events with 
$80~\GeV{}<m_{\ell\ell}<100~\GeV$. The purity of the charge-flip background is estimated to be around $80\%$ in this region.
The modelling of the $WZ+$jets background is checked in a $WZ+2$-jets region requiring the presence of an additional lepton. 
The \pt of this third lepton is shown at the bottom left of Figure~\ref{fig:Val}. The purity of the $WZ$+jets is estimated to be around $60\%$ in this region.
The modelling of backgrounds from non-prompt leptons is tested in a $b-$tagged region that requires at least one $b-$jet. 
The purity of the non-prompt lepton background is estimated to be around $80\%$ in this region.
The $m_{jj}$ modelling is checked by examining events with masses $m_{jj}$ in the regions $m_{jj}<65~\GeV$ or $m_{jj}>105~\GeV$.
The distribution of $m_{jj}$ in this region is shown at the bottom right of Figure~\ref{fig:Val}.
Finally, conversion and prompt backgrounds are tested in a region with at most one jet, called the $\le 1$ jet region. 
The purity of the conversion and prompt backgrounds is estimated to be around $70\%$ in this region.
As for the \3l\ channel, good agreement is observed between the data and the prediction in all five VRs.

\begin{figure}[h!]
\begin{center}
\includegraphics[width=0.49\textwidth]{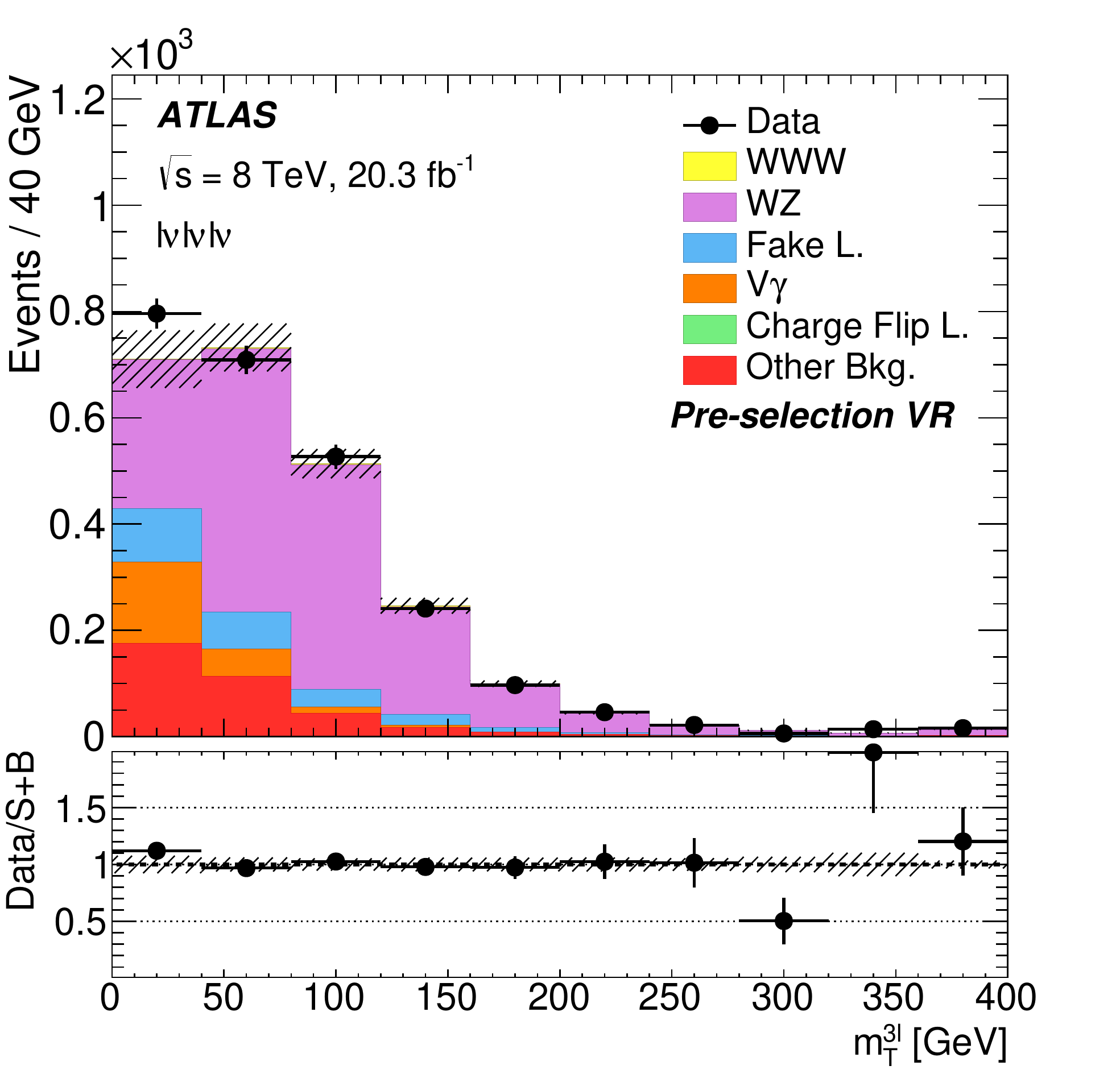}
\includegraphics[width=0.49\textwidth]{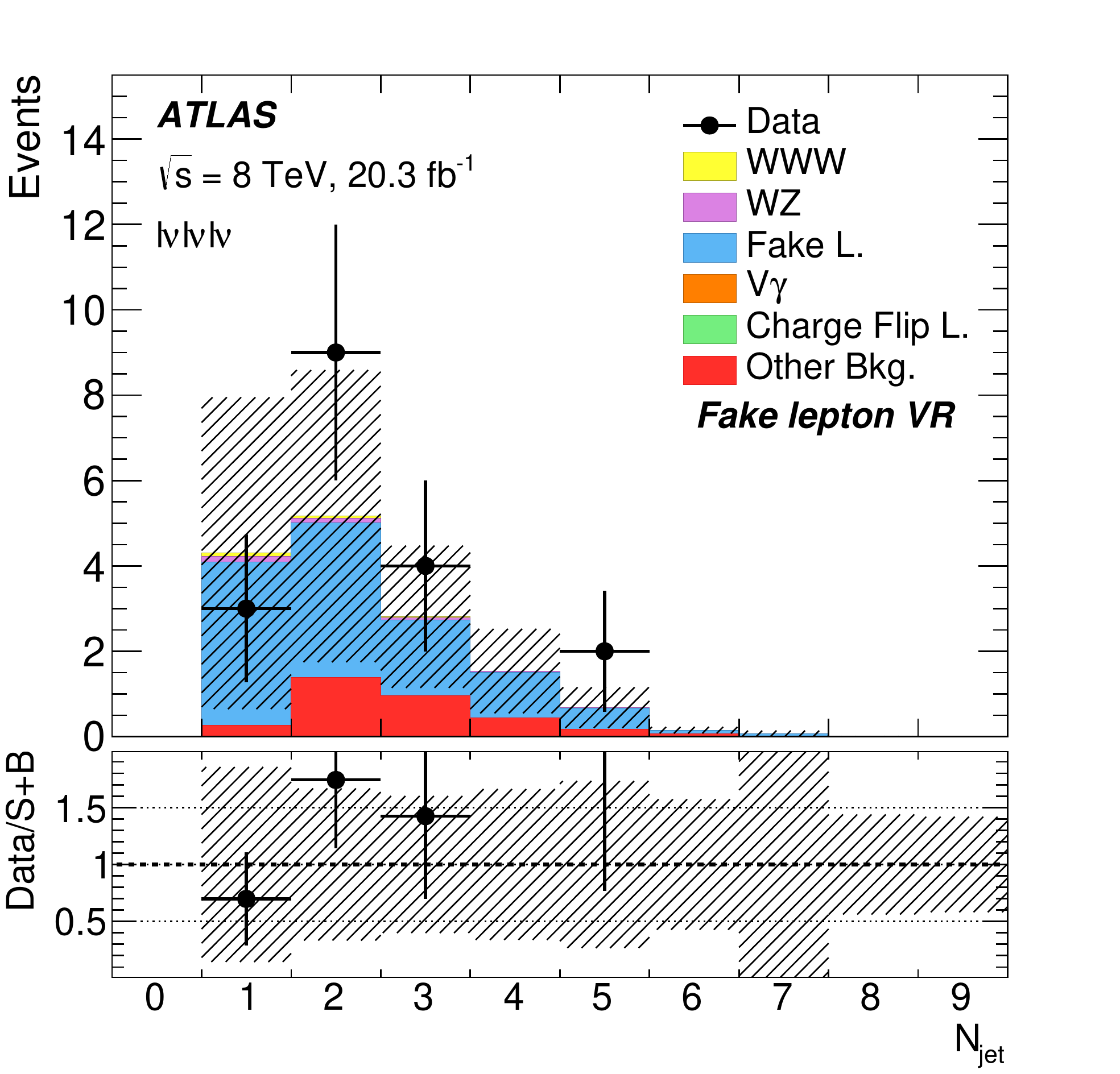}
\includegraphics[width=0.49\textwidth]{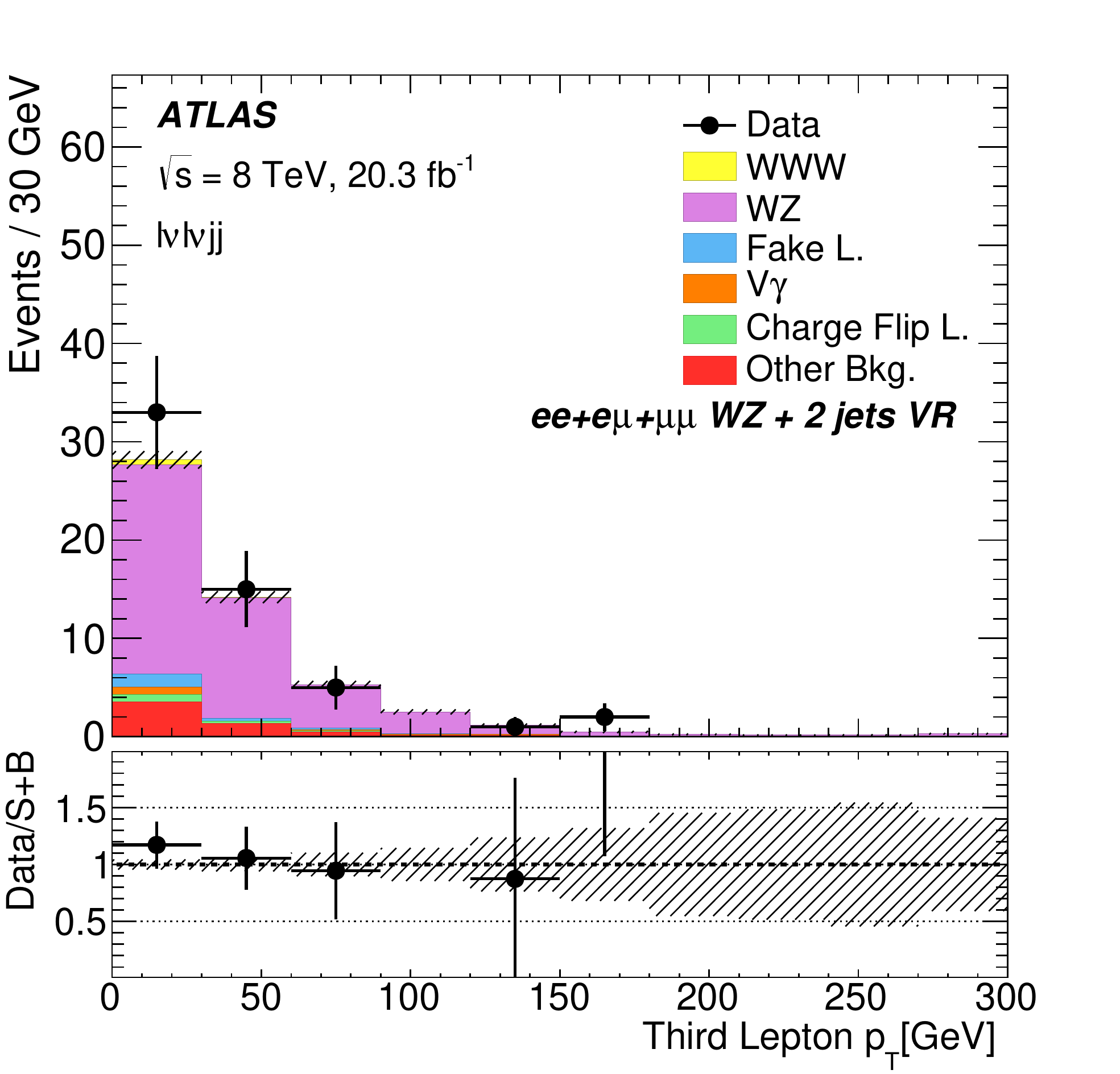} 
\includegraphics[width=0.49\textwidth]{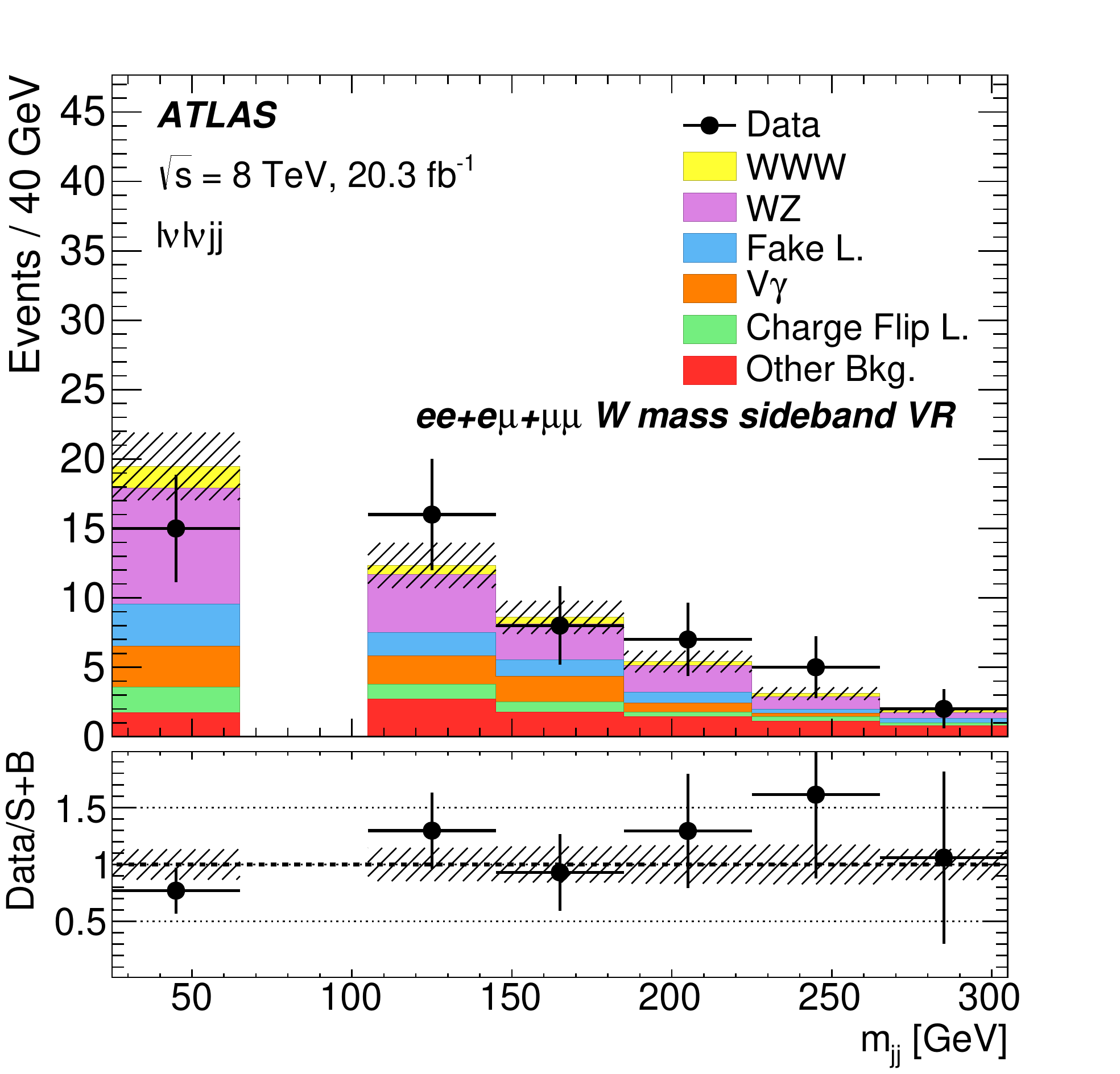} 

\end{center}
\caption{Distributions in four different VRs, two corresponding to the \3l\ channel (top)
and two to the \2l2j\ channel (bottom).
For the \3l\ channel the $m_T^{3\ell}$ distribution in the preselection region (top left)
and the jet multiplicity distribution in the fake-lepton region (top right) are shown.
For the \2l2j\ channel the third-lepton \pt in the $WZ+2$-jets region (bottom left) and 
the $m_{jj}$ distribution in the $W$ mass sideband region (bottom right) are shown.
The ``other backgrounds'' contain prompt leptons and are estimated from MC.
The hashed band represents total uncertainties on the signal-plus-background prediction.
The highest bin also includes events falling out of the range shown.}
\label{fig:Val} 
\end{figure}

\begin{table*}
	\caption{Expected numbers of signal and background events in the VRs compared
	to the numbers of events observed in data. The first uncertainty is statistical and the second is systematic.}
\begin{center}
\begin{tabular}{cl||c|c|c}
 \hline
 & Validation Region & Signal & Background & Observed \\
 \hline
 \hline
\multirow{3}{*}{\3l\ } &Preselection & $9.78 \pm 0.04 \pm 0.45$ & $2392 \pm 7 \pm 298$ & 2472   \\
\cline{2-5}
&Fake-lepton & $0.15 \pm 0.01 \pm 0.02 $  & $15 \ \ \pm 1 \pm 10$ & 18  \\
\cline{2-5}
 &$Z\gamma$ & $0.32 \pm 0.01 \pm 0.02$ & $119 \pm 3 \pm 20$ & 119  \\
\hline \hline
\multirow{6}{*}{\2l2j\ } &Charge-flip & $0.98 \pm 0.04 \pm 0.06$ & $21 \pm 1 \pm 2$ & 22 \\
\cline{2-5}
 &$WZ+2$-jets & $0.55 \pm 0.03 \pm 0.04$ & $52 \ \ \pm 1 \pm 10$ & 56  \\
\cline{2-5}
 &$b$-tagged & $1.00 \pm 0.05 \pm 0.07$ & $69 \ \ \pm 1 \pm 23$ & 78 \\
\cline{2-5}
 &$W$ mass sideband & $3.35 \pm 0.08 \pm 0.43$  & $ 48 \pm 2 \pm 6$ & 53 \\
\cline{2-5}
 &$\leq 1$ jet & $1.62 \pm 0.06 \pm 0.40$ & $139 \pm 3 \pm 18$  &  145\\
 \hline
 \end{tabular}
\end{center}
\label{tab:CR}
\end{table*}

\section{Systematic uncertainties}
Systematic uncertainties in the signal and background predictions arise from the measurement of the integrated luminosity, from the
experimental and theoretical modelling of the signal acceptance and detection efficiency, and from the background estimation. 
The effect of the systematic uncertainties on the overall signal and background yields are evaluated separately for the \3l\ and \2l2j\ channels . The results are summarised in Table~\ref{tab:syst}. The systematic uncertainties are included as nuisance parameters in the profile likelihood described in Section~\ref{sec:xsec}. Correlations of systematic uncertainties arising from common sources are maintained across signal and background processes and channels.

\begin{table*}
\caption{The effect of the various systematic uncertainties on the total signal and background yields (in percent) for both channels.}
\begin{center}
\begin{tabular}{l||cc|cc}
\hline
& \multicolumn{2}{c|}{\3l} & \multicolumn{2}{c}{\2l2j} \\
Source of Uncertainty & Signal $[\%]$ & Background $[\%]$& Signal $[\%]$& Background $[\%]$\\ \hline
\hline
Lepton ID, \et/\pt~scale and resolution& 1.6 & 1.8& 2.1 & 3.3 \\\hline
\met~modelling & 1.1 & 1.4 &  0.7 & 1.8 \\\hline
$b$-jet identification & 0.3 & 0.3 & 2.2 & 2.2  \\\hline
Jet \et~scale and resolution  & 2.3 & 2.8 & 21 & 15 \\\hline
Fake-lepton background  & 0 & 13 & 0 & 8\\\hline
Charge-flip background  & 0 & 0.04 & 0 & 2.2\\\hline
Luminosity  & 1.9 & 1.6 & 1.9 & 1.4 \\\hline
Pile-up estimate & 1.1 & 0.6 & 0.6 & 1.6 \\\hline
Trigger efficiency & 0.1 & 0.1 & 0.1 & 0.01 \\\hline
Normalization factor & 3.8 & 8 & 6.0 & 13\\\hline
Statistical & 1.2 & 3.2 & 2.7 & 5.1 \\\hline
\end{tabular}
\end{center}
\label{tab:syst}
\end{table*}

The experimental uncertainties include the uncertainties on the lepton and jet energy and momentum scales and resolutions, on the 
efficiencies of the lepton and jet reconstruction and identification, and on the modelling of \met\ and $b$-jets.
They are evaluated separately for both the signal and background estimations.
For the expected signal yield, the major contributions in the \3l\ channel come from
uncertainties in the lepton reconstruction and identification efficiencies as well as lepton energy/momentum 
resolution and scale modelling ($\pm1.6\%$), \met\ modelling ($\pm1.1\%$), and jet energy scale and resolution ($\pm2.3\%$).
The contributions in the \2l2j\ channel come from uncertainties in the lepton efficiencies and 
energy/momentum modelling ($\pm2.1\%$), \met\ modelling ($\pm0.7\%$), $b$-jet identification ($\pm2.2\%$), and jet energy resolution and scale modelling ($\pm21\%$).
Larger systematic uncertainties due to the jet energy scale and resolution are expected in the \2l2j\ channel due to the dijet requirements, in particular in the dijet invariant mass.
For the background yields estimated from MC simulation, the major contributions in the \3l\ channel come from 
uncertainties in lepton reconstruction and identification efficiencies ($\pm1.8\%$), \met\ modelling ($\pm1.4\%$), and jet energy modelling ($\pm2.8\%$). 
The major contributions in the \2l2j\ channel come from 
uncertainties in lepton efficiencies and energy modelling ($\pm3.3\%$), \met\ modelling ($\pm1.8\%$), $b$-jet identification ($\pm2.2\%$), and jet energy modelling ($\pm15\%$).

The estimates of the data-driven fake-lepton background also have uncertainties specific to each channel.
In the \3l\ channel, the systematic uncertainty results from the uncertainties
on the probabilities of candidate leptons that satisfy the looser lepton selection criteria to also satisfy the signal lepton selection criteria. 
For prompt leptons this uncertainty is $\pm$(5--10)\% while for fake leptons and misidentified/non-prompt leptons this uncertainty 
is $\pm$(80--90)\%. The latter uncertainty is a conservative estimate which accounts for differences in the
heavy-flavour and light-flavour composition between the signal region and the control region where the fake-lepton efficiency is determined for these leptons.
In the \2l2j\ channel, the systematic uncertainty results from the uncertainties in the measurement of the fake factors, which is estimated to be $\pm$(20--30)\%. 
Statistical uncertainties in the samples used for the matrix method and the fake-factor method also contribute to the 
overall uncertainty of the estimation of the fake-lepton background.  
The total uncertainty in the overall fake-lepton background yield is $\pm13\%$ in the \3l\ channel and $\pm8\%$ in the \2l2j\ channel. 

The charge-flip background is only relevant for the \ssee/\ssem\ final state in the \2l2j\ channel and 
for the 0-SFOS region in the \3l\ channel. Its uncertainty is
dominated by the statistical precision with which the electron charge misidentification rate is determined from the available data. 
Since the charge-flip background estimation uses the number of $Z(\rightarrow e^+ e^-)+2$ jets events, 
the number of events in the data also contributes to the overall systematic uncertainty. 
In the \3l\ channel, the uncertainty on the charge-flip background estimate is $\pm0.5\%$ in the 0-SFOS region
but is $\pm0.04\%$ for the total background estimate in all three signal regions. 
In the \2l2j\ channel, the total systematic uncertainty in the 
overall background yield due to the uncertainty in the charge-flip background estimate is found to be $\pm2.2\%$.

There are also uncertainties in the overall normalization of the signal and MC background
cross sections. Uncertainties in the signal cross section are those described in Section~\ref{sec:signal}. These are not, however, included as uncertainties in the model and merely
serve as a comparison for the final measurement in Section~\ref{sec:xsec}.
The normalizations of the SM background cross sections described in Section~\ref{sec:bg_est} have their own associated uncertainties.
The uncertainty in the predicted $WZ$+jets background cross section is the most important one since it is the largest irreducible
background. The size of the uncertainty relative to the predicted $WZ$+jets background is $\pm10\%$ in 
the \3l\ channel and $\pm$(16--23)\% in the \2l2j\ channel depending on the production mechanism. This uncertainty is based on the measurement performed in the control region, for the \3l\ channel as described in Section~\ref{sec:bg}, while it is a combination of the scale, PDF and parton shower uncertainties estimated as in Ref.~\cite{atlas5}, for the \2l2j\ channel. The remaining uncertainties
are mostly negligible in the overall background prediction. The normalization uncertainty in the total background
prediction is around $\pm 8$\% in the \3l\ channel and $\pm 13$\% in the \2l2j\ channel.

The uncertainty on the integrated luminosity is $\pm1.9\%$, affecting the overall normalization of
both the signal and background processes estimated from MC simulation. It is derived following the methodology detailed in Ref.~\cite{lumi2}.
The uncertainties associated with the pile-up reweighting of the events are estimated to be no more than $\pm0.1\%$ for the signal and the backgrounds.

\section{Cross-section measurement}
\label{sec:xsec}
The signal and background predictions together with their uncertainties are compared to the data for six signal regions in Table~\ref{tab:yield}.
The expected signal yields are calculated using the SM \www\ cross sections listed in Section~\ref{sec:signal}. 
The expected numbers of signal plus background events are consistent with the numbers of events observed in data in all regions. 
Figure~\ref{fig:Sig} shows the $m_{T}^{3\ell}$ 
distribution for the \3l\ channel and the distribution of the sum of the scalar \pt\ for all selected objects, $\Sigma \pt = \pt^{\ell,1}+\pt^{\ell,2}+\pt^{j,1}+\pt^{j,2}+\met$, for the \2l2j\ channel, after summing over the three signal regions in each channel. 
Good agreement between data and the signal-plus-background model is observed for both distributions. 

\begin{table*}
\caption{Numbers of expected signal and background events, and their statistical and systematic uncertainties, together with the observed yields in the data in the signal regions for the two channels. 
}
\label{tab:yield}
\begin{center}
\small
\renewcommand{\tabcolsep}{3pt}
\begin{tabular}{l||lclcl|lclcl|lclcl}
\hline
\multicolumn{1}{c||}{\3l} & \multicolumn{5}{c|}{0~SFOS} & \multicolumn{5}{c|}{1~SFOS} & \multicolumn{5}{c}{2~SFOS} 
\Tstrut
\\[0.2cm]
\hline
\multicolumn{1}{c||}{\www\ signal} &  $1.34 $&$\pm$&$ 0.02$&$\pm$&$0.07$ &  $1.39 $&$\pm$&$ 0.02 $&$\pm$&$ 0.08 $ &  $0.61 $&$\pm$&$ 0.01 $&$\pm$&$ 0.03$ 
\Tstrut
\\[0.2cm]
\hline
$WZ$ &  $0.59 $&$\pm$&$ 0.00$&$\pm$&$ 0.07$ &  $11.9 $&$\pm$&$ 0.1 $&$\pm$&$ 1.3$ &  $9.1 $&$\pm$&$ 0.1 $&$\pm$&$ 1.0$ \Tstrut \\[0.2cm]
Other prompt background&  $0.21 $&$\pm$&$ 0.01 $&$\pm$&$ 0.02$ &  $0.78 $&$\pm$&$ 0.02 $&$\pm$&$ 0.11$ &  $0.60 $&$\pm$&$ 0.02 $&$\pm$&$ 0.10$ \\[0.2cm]
Charge-flip background&  $0.04 $&$\pm$&$ 0.00 $&$\pm$&$ 0.01$ &  \multicolumn{5}{c|}{-} &  \multicolumn{5}{c}{-} \\[0.2cm]
$V\gamma$ &  \multicolumn{5}{c|}{-} &  $0.20 $&$\pm$&$ 0.13 $&$\pm$&$ 0.29$ &  $0.11 $&$\pm$&$ 0.10 $&$\pm$&$ 0.29$ \\[0.2cm]
Fake-lepton background&  $1.5 $&$\pm$&$ 0.3 $&$\pm$&$ 1.4$ &  $1.9 $&$\pm$&$ 0.3 $&$\pm$&$ 1.9$ &  $0.49 $&$\pm$&$ 0.16 $&$\pm$&$ 0.47$ \\[0.2cm]
\multicolumn{1}{l||}{Total background} &  $2.4 $&$\pm$&$ 0.3 $&$\pm$&$ 1.4 $ &  $14.8 $&$\pm$&$ 0.4 $&$\pm$&$ 2.3$ &  $10.3 $&$\pm$&$ 0.2 $&$\pm$&$ 1.2$ \Tstrut \\[0.2cm]
\hline
\multicolumn{1}{l||}{Signal $+$ background} &  $3.7 $&$\pm$&$ 0.3 $&$\pm$&$ 1.4 $ &  $16.2 $&$\pm$&$ 0.4 $&$\pm$&$ 2.3 $ &  $10.9 $&$\pm$&$ 0.2 $&$\pm$&$ 1.2$ \Tstrut \\[0.2cm]
\multicolumn{1}{l||}{Data} &  \multicolumn{5}{c|}{$5$} &  \multicolumn{5}{c|}{$13$} &  \multicolumn{5}{c}{$6$} \\[0.2cm]
\hline
\hline
\multicolumn{1}{l||}{\2l2j} & \multicolumn{5}{c|}{\ssee} & \multicolumn{5}{c|}{\ssem} & \multicolumn{5}{c}{\ssmm} \Tstrut \\[0.2cm]
\hline
\multicolumn{1}{l||}{\www\ signal} & 0.46 &$\pm$& 0.03 &$\pm$&$ 0.07$  & 1.35 &$\pm$& 0.05 &$\pm$&$ 0.19$  & 1.65 &$\pm$& 0.06 &$\pm$&$ 0.30$ \Tstrut \\[0.2cm] 
\hline
$WZ$ & 0.74 &$\pm$&  0.13 &$\pm$&$ 0.44$  & 2.77 &$\pm$&  0.27 &$\pm$&$ 0.66$  & 3.28 &$\pm$&  0.29 &$\pm$&$ 0.71$ \Tstrut \\[0.2cm] 
Other prompt background& 0.46 &$\pm$&  0.05 & $\pm$ & 0.16  & 1.33 &$\pm$&  0.10 &$\pm$&$ 0.38$  & 1.33 &$\pm$&  0.15 &$\pm$&$ 0.38$ \\[0.2cm] 
Charge-flip background& 1.13 &$\pm$&  0.13 &$\pm$&$ 0.24$  & 0.74 &$\pm$&  0.08 &$\pm$&$ 0.16$  & \multicolumn{5}{c}{-} \\[0.2cm] 
$V\gamma$ & 0.75 &$\pm$&  0.35 &$\pm$&$ 0.21 $  & 2.5 &$\pm$&  0.7 &$\pm$&$ 0.7$  & \multicolumn{5}{c}{-} \\[0.2cm] 
Fake-lepton background& 0.96 &$\pm$&  0.15 &$\pm$&$ 0.39$  & 2.04 &$\pm$&  0.22 &$\pm$&$ 0.89$  & 0.43 &$\pm$&  0.06 &$\pm$&$ 0.25$ \\[0.2cm] 
\multicolumn{1}{l||}{Total background} & 4.0 &$\pm$&  0.4 &$\pm$&$ 0.7$  & 9.4 &$\pm$&  0.8 &$\pm$&$ 1.4$  & 5.0 &$\pm$&  0.3 &$\pm$&$ 0.8$ \Tstrut \\[0.2cm]
\hline
\multicolumn{1}{l||}{Signal $+$  background} & 4.5 &$\pm$& 0.4 &$\pm$&$ 0.7$  & 10.7 &$\pm$& 0.8 &$\pm$&$ 1.4$  & 6.7 &$\pm$& 0.3 &$\pm$&$ 0.9$ \Tstrut \\[0.2cm]
\multicolumn{1}{l||}{Data} & \multicolumn{5}{c|}{0} & \multicolumn{5}{c|}{15} & \multicolumn{5}{c}{6}\\[0.2cm]
\hline
\end{tabular}
\end{center}
\end{table*}

The amount of \www\ signal in the selected data set is determined using the numbers of expected signal and background events as well the numbers of observed events in the data. 
The signal strength, $\mu$, is the parameter of interest, defined as a scale factor multiplying the cross section times 
branching ratio predicted by the SM. 
A test statistic based on the profile-likelihood ratio~\cite{profile} is 
used to extract $\mu$ from a maximum-likelihood fit of the signal-plus-background model to the data. 
The likelihood, $\cal{L}$, is given by
\begin{equation}
{\cal{L}} = \prod_c \prod_i {\text{Poisson}} \left[  n_{i_c}^{\text{obs}} \mid \mu \times n_{i_c}^{\text{sig, SM}} (\theta_k) + n_{i_c}^{\text{bkg}}(\theta_k) \right] \prod_k g(\theta_k)
\end{equation}
where the index $c$ represents one of the two analysis 
channels, $i$ represents one of the three signal regions in each channel, $n^{\text{obs}}$ is number of observed events, 
$n^{\text{sig, SM}}$ is the expected number of signal events based on 
the SM calculations, and $n^{\text{bkg}}$ is the expected number of background events. 
The effect of a systematic uncertainty $k$ on the likelihood is 
modelled with a nuisance parameter, $\theta_k$, constrained with a corresponding 
Gaussian probability density function $g(\theta_k)$. 

The test statistic, $t_\mu$, is defined as 
\begin{equation}
t_{\mu} = -2 \ln \lambda(\mu)=-2 \ln \frac{{\cal{L}}(\mu, \hat{\hat{\theta}}(\mu))}{{\cal{L}}(\hat{\mu}, \hat{\theta})} 
\end{equation}
where $\hat{\mu}$ is the unconditional maximum-likelihood (ML) estimators of the independent signal strength in the categories being compared, 
$\hat{\theta}$ are the unconditional ML estimators of the nuisance parameters, and $\hat{\hat{\theta}}(\mu)$ are the conditional ML estimators of $\theta$ for a given value of $\mu$. 
The significance of $\mu$ is obtained with the above test statistic, and is estimated using 100,000 MC pseudo-experiments to determine 
how well the fit result agrees with the background-only hypothesis. 
The observed (expected) significance of a positive signal cross section is $0.96~\sigma$ 
($1.05~\sigma$) for the combination of the two channels.
Most of the sensitivity comes from the 0-SFOS category in the \3l\ channel and the \ssmm\ category in the \2l2j\ channel.
The most significant deviation from the signal-plus-background hypothesis
occurs in the $e^\pm e^\pm$ region, where zero events are observed and 4.0 background and 0.46 signal events are expected. 
The probability that the background fluctuates down to zero events is 2.3\%.

The central value of $\mu$ corresponds to the minimum of the negative log-likelihood distribution. 
The measured fiducial cross section in each channel is obtained by multiplying $\mu$ by the expected 
value of the fiducial cross section in that channel. The measured total cross section is obtained by combining the results for the
two channels and then extrapolating to the total phase space using the signal acceptance. 
The log-likelihood scans for the total cross-section measurement are evaluated with 
and without systematic uncertainties and are shown in Figure~\ref{fig:Sig-likelihood}. 
The expected and observed fiducial and total cross sections are summarized in Table~\ref{tab:measurements}.

The presence of the \www\ signal is also assessed using a 
one-sided 95\% CL upper limit on the production cross section using the 
CL$_s$ method of Ref.~\cite{CLsMethod}. The limits are evaluated using 2,000 MC pseudo-experiments.
The observed (expected) upper limit on the fiducial cross section in the absence of \www\ production
is found to be 1.3~fb (1.1~fb) in the \3l\ channel and 1.1~fb (0.9~fb) in the \2l2j\ channel. The observed (expected) upper limit in the absence of \www\ production on the total cross section is 730~fb (560~fb) when the two channels are combined. 
If the SM \www\ signal is also considered, the expected upper limit on the total cross section is 850~fb.

\begin{table*}
\caption{
The predicted and observed fiducial cross sections for the \3l\ and \2l2j\ channels and the predicted and observed total cross section for the combination of the two channels.
}
\begin{center}

\begin{tabular}{rc||c|c|c}
\hline
	   && \multicolumn{2}{c}{Cross section [fb]} \\
           && Theory & \multicolumn{1}{c}{Observed}\\
\hline \hline
\multirow{2}{*}{Fiducial}& \3l &  
$0.309 \pm 0.007~({\rm stat.})~\pm 0.015~({\rm PDF}) \pm 0.008~({\rm scale})$&
\multicolumn{1}{c}{$0.31~^{+0.35}_{-0.33}$~(stat.)~$^{+0.32}_{-0.35}$~(syst.)}
\Tstrut 
\\[4pt]
\cline{2-4}
& \2l2j & 
$0.286 \pm 0.006~({\rm stat.})~\pm 0.015~({\rm PDF}) \pm 0.010~({\rm scale})$&

\multicolumn{1}{c}{$0.24~^{+0.39}_{-0.33}$~(stat.)~$^{+0.19}_{-0.19}$~(syst.)}
\Tstrut 
\\[4pt]
\hline
\multicolumn{2}{c||}{Total} & 
$241.5 \pm 0.1$~(stat.)~$\pm~10.3$~(PDF) $\pm~6.3$~(scale) 

&
\multicolumn{1}{c}{$230~{\pm 200}$~(stat.)~$^{+150}_{-160}$~(syst.)}
\Tstrut 
\\[4pt]
\hline
\end{tabular}
\end{center}
\label{tab:measurements}
\end{table*}

\begin{figure}[h!]
\begin{center}
\includegraphics[width=.49\textwidth]{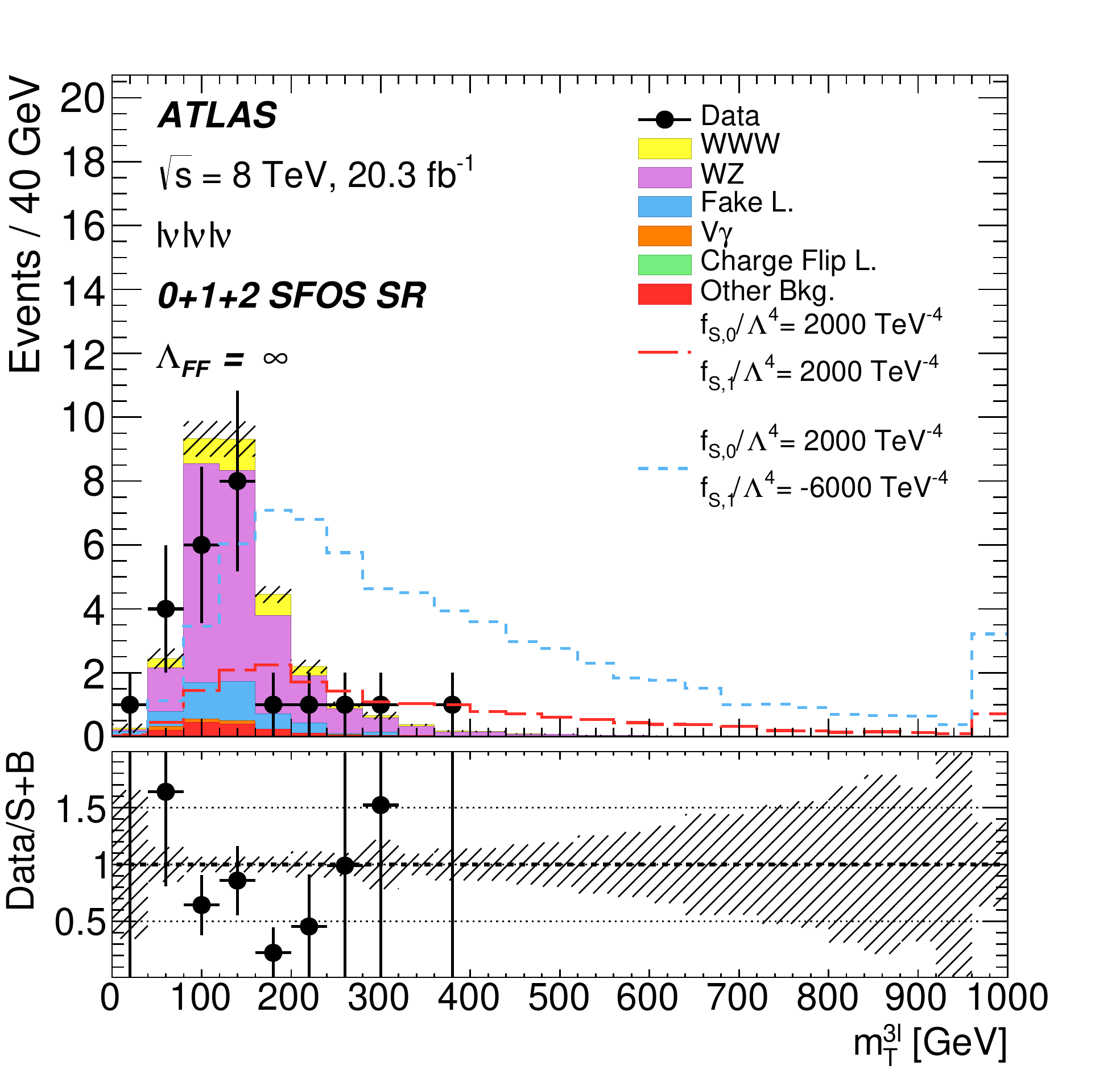}
\includegraphics[width=.49\textwidth]{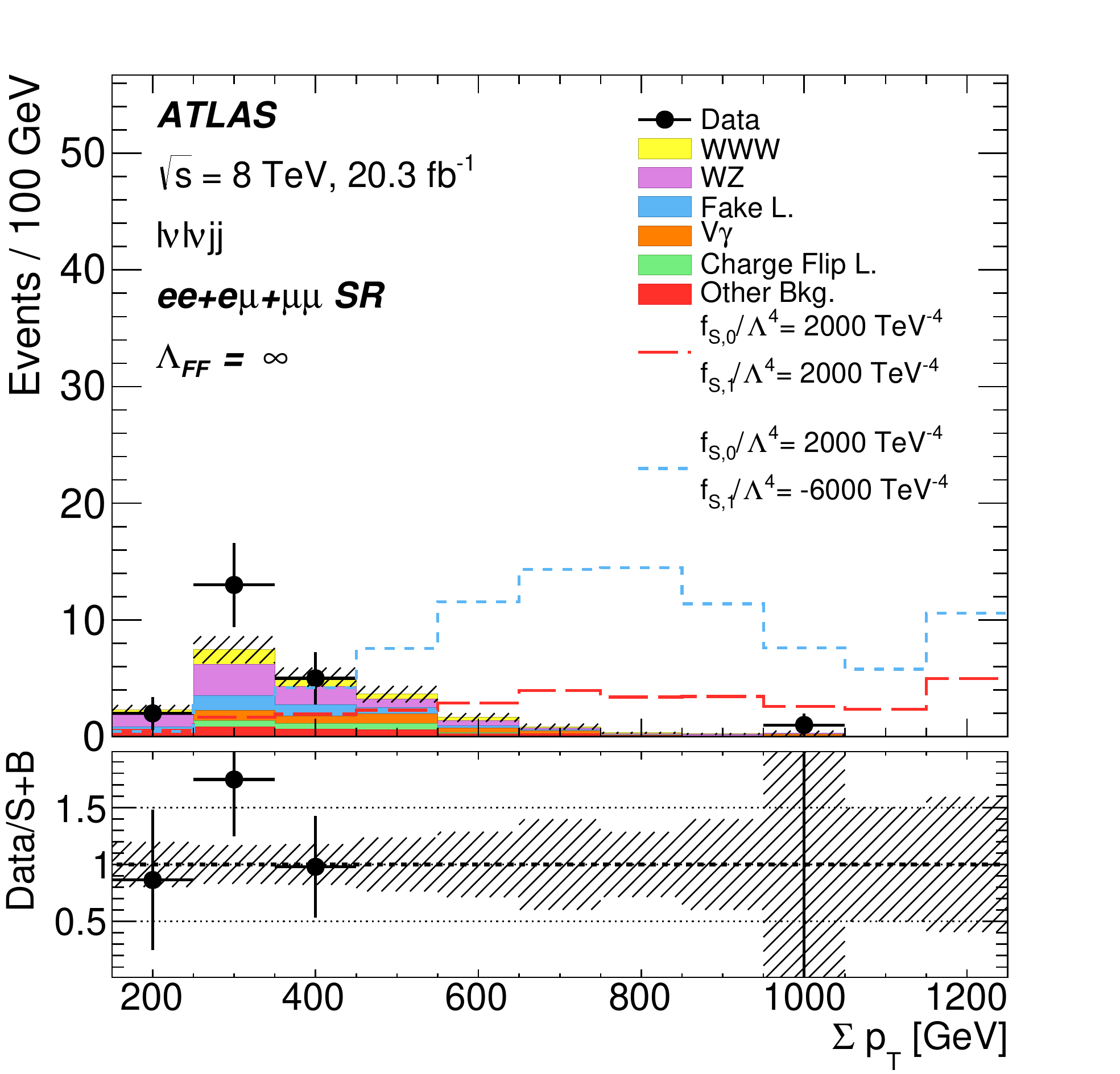} 
\end{center}
\caption{The distribution of $m_T^{3\ell}$ for the \3l\ channel (left) and the distribution of $\Sigma \pt$ for the \2l2j\ channel (right) as observed in the data (dots with error bars indicating the statistical uncertainties) and as expected from SM signal and background processes.
The ratios between the observed numbers of events in data and the expected SM signal plus background contributions are shown in the lower panels. The hashed bands results from the systematic uncertainties on the sum of the signal plus background contributions. The ``other backgrounds'' contain prompt leptons and are estimated from MC.
Contributions from aQGCs are also shown, assuming the non-unitarized case ($\Lambda_{\rm FF} = \infty$) and two different sets of \fszero{} and \fsone{} configurations ($\fszero{}=2000~\TeV{}^{-4}$, $\fsone{}=2000~\TeV{}^{-4}$ and $\fszero{}=2000~\TeV{}^{-4}$, $\fsone{}=-6000~\TeV{}^{-4}$). The highest bin also includes events falling out of the range shown.}
\label{fig:Sig} 
\end{figure}

\begin{figure}[h!]
\begin{center}
\includegraphics[width=0.5\textwidth]{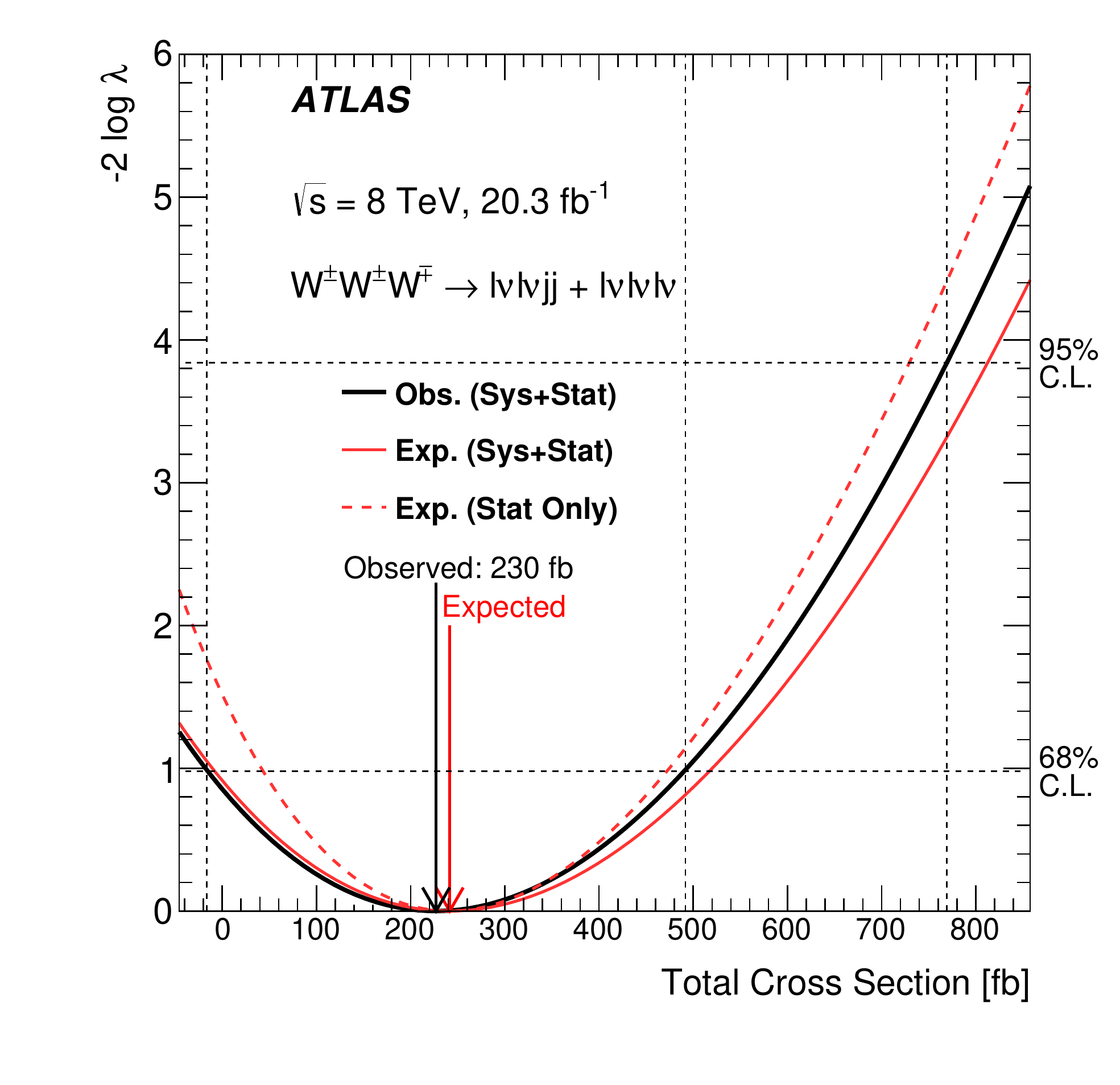} 
\end{center}
\caption{Profile-likelihood scans as a function of the total cross section for the combination of all six signal regions. The expected (red) scans are shown when considering only statistical uncertainties (dashed line) and when considering both statistical and systematic uncertainties (solid line). The observed (black solid line) scan is also shown. The dotted black grid-lines pinpoint the location of the 68\% and 95\% CL uncertainties in the measurement of the signal strength.}
\label{fig:Sig-likelihood} 
\end{figure}

\section{Limits on anomalous quartic gauge couplings (aQGCs)}
Contributions from sources beyond the SM to the \www\ production process can be expressed in a model-independent way using higher-dimensional operators leading to $WWWW$ aQGCs. 
The parameterization of aQGCs is based on Ref.~\cite{aQGC} in a linear representation~\cite{representation} considering only dimension-eight operators involving four gauge bosons.
There are 18 dimension-eight operators built from the covariant derivative of the Higgs field $D_\mu \Phi$, the SU(2)$_{\textrm{L}}$ field strength $W^i_{\mu\nu}$, 
and U(1)$_{\textrm{Y}}$ field strength $B_{\mu\nu}$. Only the two terms built exclusively from $D_\mu \Phi$ and with aQGC parameters \fszero{} and \fsone{} are considered in this analysis:
\begin{equation}
 {\cal{L}}_{S,0} = \frac{f_{S,0}}{\Lambda^4} [(D_\mu\Phi)^\dagger D_\nu \Phi] \times [(D^\mu\Phi)^\dagger D^\nu \Phi],
 \label{eq:EffLag1}
\end{equation}
\begin{equation}
 {\cal{L}}_{S,1} = \frac{f_{S,1}}{\Lambda^4} [(D_\mu\Phi)^\dagger D^\mu \Phi] \times [(D_\nu\Phi)^\dagger D^\nu \Phi],
  \label{eq:EffLag2}
 \end{equation}
where $\Lambda$ is the energy scale of the new physics. These two operators only affect massive bosons and do not depend on the gauge boson 
momenta since no SU(2)$_{\textrm{L}}$ or U(1)$_{\textrm{Y}}$ field strengths are included. As a result, they are important for the study of longitudinal vector-boson scattering. 
Similar parameters were studied before by the ATLAS and CMS Collaborations in Refs.~\cite{atlas1,atlas3,cms2} 

The effective Lagrangian approach leads to tree-level unitarity violation. This can be avoided by introducing a form factor~\cite{formfactor} as
\begin{equation}
\alpha \rightarrow \frac{\alpha_0}{(1+\hat{s}/\Lambda_{\rm FF}^2)}
\end{equation}
where $\alpha$ corresponds to one of the two couplings, $\alpha_0$ is the value of the aQGC at low energy, 
$\hat{s}$ is the square of the partonic centre-of-mass energy, and $\Lambda_{\rm FF}$ is the form-factor cutoff scale. 
However, there is no theoretical algorithm to predict for which form-factor cutoff scale the cross section would violate unitarity. 
Therefore different values of $\Lambda_{\rm FF}$ are considered with $\Lambda_{\rm FF}=$ 0.5, 1, 2, and 3~\TeV{} as well 
as $\Lambda_{\rm FF} = \infty$, which corresponds to the non-unitarized case.  

Events with aQGCs are generated with {\sc vbfnlo} at LO and passed through the ATLAS detector simulation. A grid of samples is obtained using different parameters of $\fszero{}$ and $\fsone{}$ values. The interpolation between these points is performed with a 2-dimensional quadratic function in the ($\fszero{}$, $\fsone{}$) space. The LO samples are scaled using a factor derived from the ratio of the SM LO to NLO predictions. Figure~\ref{fig:Sig} show the expected distribution for the non-unitarized ($\Lambda_{\rm FF} = \infty$) aQGC signal samples being generated with parameters $\fszero{}=2000~\TeV{}^{-4}$, $\fsone{}=2000~\TeV{}^{-4}$ in red and parameters $\fszero{}=2000~\TeV{}^{-4}$, $\fsone{}=-6000~\TeV{}^{-4}$ in blue as a function of the $m_T^{3\ell}$ distribution in the \3l\ channel and the $\Sigma \pt$ distribution in the \2l2j\ channel, summed over the three signal regions in each channel.
Even though aQGC events tend to have leptons or jets with larger momenta, the detection efficiency for events in the fiducial region is found to be consistent with the one obtained for the SM sample within 20\%. The efficiencies of the aQGC samples are used with their statistical and systematic uncertainties to derive the 95\% confidence intervals (CI) on aQGC, while the largest observed deviation of the aQGC efficiencies from the SM one is used as an extra systematic uncertainty.
Frequentist CI on the anomalous coupling are computed by forming a 
profile-likelihood-ratio test that incorporates the observed and expected numbers of signal events for different values of the anomalous couplings.
Table~\ref{tab:aQGC} shows the expected and observed 95\% CI on \fszero~(\fsone) with different $\Lambda_{\rm FF}$ values, assuming \fsone~(\fszero) to be zero. 
Figure~\ref{fig:Sig-aQGC} shows the two-dimensional 95\% CL contour limits of \fszero~vs \fsone{} in the cases where $\Lambda_{\rm FF} = 1$~\TeV{} and
$\Lambda_{\rm FF} = \infty$.
For $\Lambda_{\rm FF} = \infty$, the limits can be compared to the stronger
limits obtained by the CMS Collaboration in Ref.~\cite{cms2} in a different production
channel. Other parameterization ($\alpha_4$, $\alpha_5$) of new physics have been introduced in Ref.~\cite{aQGC_a4a5_1,aQGC_a4a5_2,aQGC_a4a5_3}.
The limits presented in this paper can be converted into limits on $\alpha_4$ 
and $\alpha_5$ following the formalism defined in the Appendix of Ref.~\cite{aQGC} 
and using Equations~(60)~and~(61) in Ref.~\cite{Degrande:2013rea}. For example, non-unitarized 
limits obtained for $\Lambda_{\rm FF} = \infty$ are: $\alpha_4$ expected [$-0.61$, $0.78$], 
$\alpha_4$ observed [$-0.49$, $0.75$] and $\alpha_5$ expected [$-0.57$,$0.69$], $\alpha_5$ observed [$-0.48$,$0.62$].
Limits derived by the ATLAS Collaboration in other final states are reported in Refs.~\cite{atlas1, atlas3}. 
The latter were obtained using a different unitarization scheme. 
Since that scheme is not applicable to triboson production, a combination of the limits is not possible.

\begin{table}[h!]
    \caption{Expected and observed 95\% CI on \fszero~(\fsone) for different $\Lambda_{\rm FF}$ values, assuming \fsone~(\fszero) to be zero.}
  \begin{center}
    \begin{tabular}{  c | c  c | c  c  }
      \hline
      $\Lambda_{\rm FF}$& \multicolumn{2}{c|}{Expected CI [$\times 10^4\,$\TeV{}$^{-4}$]} &\multicolumn{2}{c}{Observed CI [$\times 10^4\,$\TeV{}$^{-4}$]} \\
      ~[\TeV{}]~ & \fszero & \fsone& \fszero & \fsone\\\hline
      $0.5$ &    [$-0.79$, 0.89] & [$-1.06$, 1.27] & [$-0.74$, 0.86] & [$-0.99$, 1.20] \\\hline
      $1$ &      [$-0.36$, 0.41] & [$-0.52$, 0.60] & [$-0.34$, 0.40] & [$-0.48$, 0.58] \\\hline
      $2$ &      [$-0.22$, 0.25] & [$-0.33$, 0.39] & [$-0.20$, 0.24] & [$-0.29$, 0.36] \\\hline
      $3$ &      [$-0.19$, 0.22] & [$-0.29$, 0.36] & [$-0.16$, 0.21] & [$-0.25$, 0.33] \\\hline
      $\infty$ & [$-0.16$, 0.19] & [$-0.25$, 0.30] & [$-0.13$, 0.18] & [$-0.21$, 0.27] \\\hline
    \end{tabular}
    \label{tab:aQGC}
  \end{center}
\end{table}

\begin{figure}[h!]
\begin{center}
\includegraphics[width=0.49\textwidth]{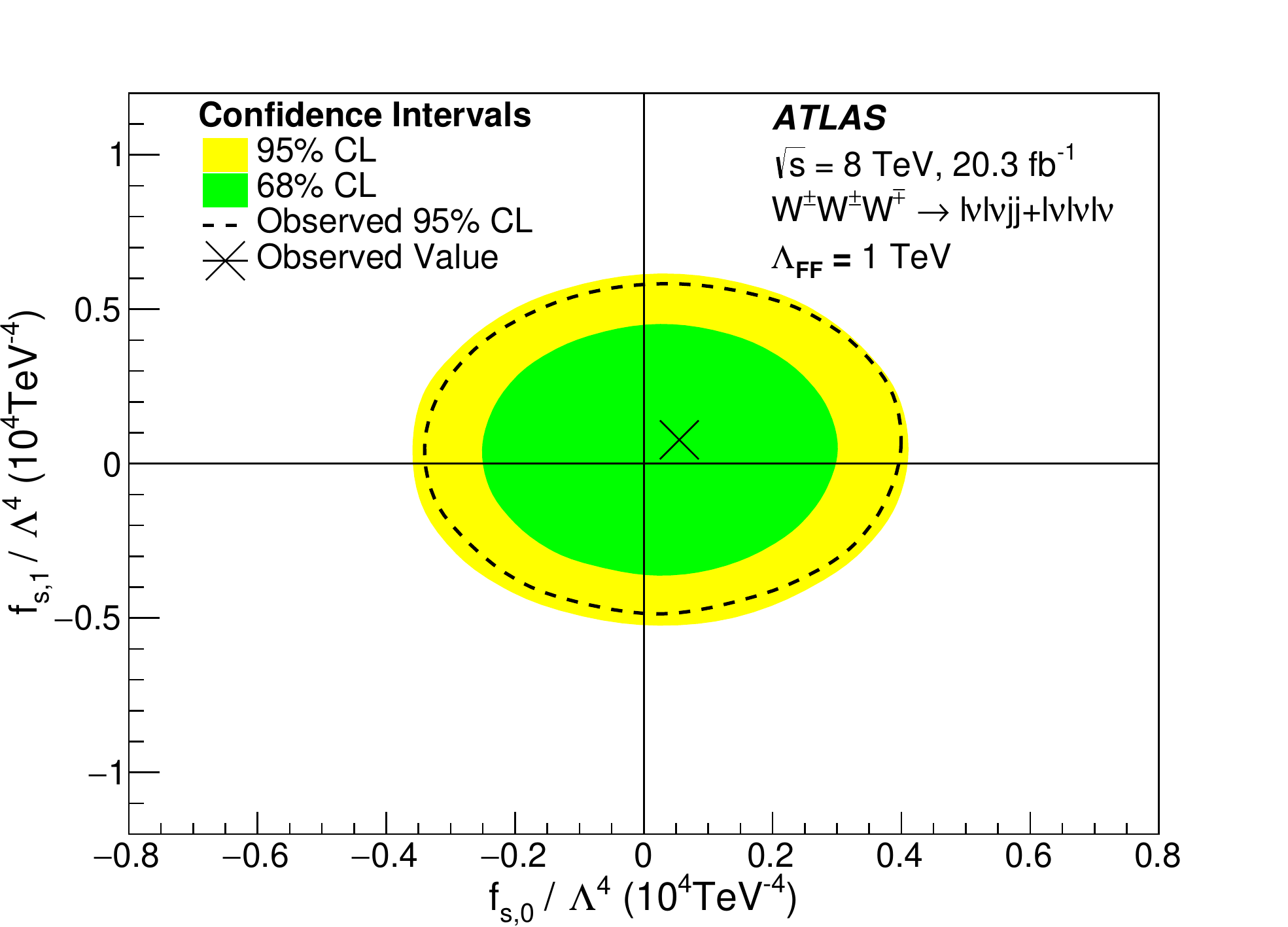} 
\includegraphics[width=0.49\textwidth]{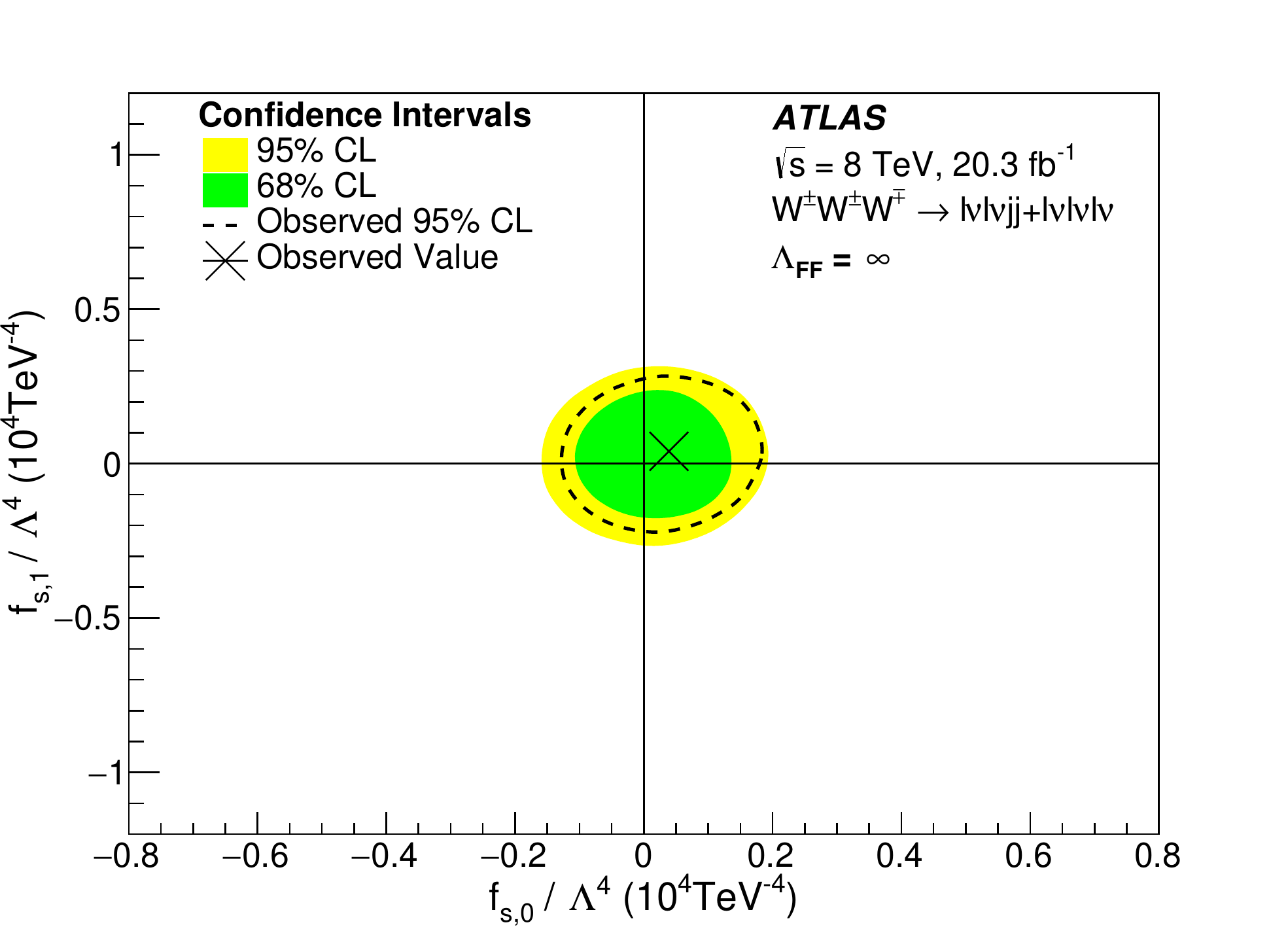} 
\end{center}
\caption{Expected 68\% and 95\% CL contours for \fsone~vs \fszero~compared to the observed 95\% CL contour and the observed best-fit value for cases when $\Lambda_{\rm FF}=1$~\TeV{} (left) and $\Lambda_{\rm FF}=\infty$ (right).}
\label{fig:Sig-aQGC} 
\end{figure}

\section{Summary}
A search for triboson \www\ production in two decay channels (${\wwwlep}$~and\\${\wwwsemilep}$~with~$\ell=e, \mu$) is reported,
using proton-proton collision data corresponding to an integrated luminosity of 20.3~$\ifb$ at a centre-of-mass energy of 8~\TeV\ collected by the ATLAS detector at the LHC. 
Events with exactly three charged leptons or two same-charge leptons in association with two jets are selected. 
The data are found to be in good agreement with the SM predictions in all signal regions. 
The observed 95\% CL upper limit
on the SM \www\ production cross section is found to be 730~fb with an expected limit of 560~fb in the absence of \www\ production.
Limits are also set on the aQGC parameters \fszero~and \fsone.

%% file: Acknowledgements.tex

We thank CERN for the very successful operation of the LHC, as well as the
support staff from our institutions without whom ATLAS could not be
operated efficiently.

We acknowledge the support of ANPCyT, Argentina; YerPhI, Armenia; ARC, Australia; BMWFW and FWF, Austria; ANAS, Azerbaijan; SSTC, Belarus; CNPq and FAPESP, Brazil; NSERC, NRC and CFI, Canada; CERN; CONICYT, Chile; CAS, MOST and NSFC, China; COLCIENCIAS, Colombia; MSMT CR, MPO CR and VSC CR, Czech Republic; DNRF and DNSRC, Denmark; IN2P3-CNRS, CEA-DSM/IRFU, France; GNSF, Georgia; BMBF, HGF, and MPG, Germany; GSRT, Greece; RGC, Hong Kong SAR, China; ISF, I-CORE and Benoziyo Center, Israel; INFN, Italy; MEXT and JSPS, Japan; CNRST, Morocco; FOM and NWO, Netherlands; RCN, Norway; MNiSW and NCN, Poland; FCT, Portugal; MNE/IFA, Romania; MES of Russia and NRC KI, Russian Federation; JINR; MESTD, Serbia; MSSR, Slovakia; ARRS and MIZ\v{S}, Slovenia; DST/NRF, South Africa; MINECO, Spain; SRC and Wallenberg Foundation, Sweden; SERI, SNSF and Cantons of Bern and Geneva, Switzerland; MOST, Taiwan; TAEK, Turkey; STFC, United Kingdom; DOE and NSF, United States of America. In addition, individual groups and members have received support from BCKDF, the Canada Council, CANARIE, CRC, Compute Canada, FQRNT, and the Ontario Innovation Trust, Canada; EPLANET, ERC, ERDF, FP7, Horizon 2020 and Marie Sk{\l}odowska-Curie Actions, European Union; Investissements d'Avenir Labex and Idex, ANR, R{\'e}gion Auvergne and Fondation Partager le Savoir, France; DFG and AvH Foundation, Germany; Herakleitos, Thales and Aristeia programmes co-financed by EU-ESF and the Greek NSRF; BSF, GIF and Minerva, Israel; BRF, Norway; CERCA Programme Generalitat de Catalunya, Generalitat Valenciana, Spain; the Royal Society and Leverhulme Trust, United Kingdom.

The crucial computing support from all WLCG partners is acknowledged gratefully, in particular from CERN, the ATLAS Tier-1 facilities at TRIUMF (Canada), NDGF (Denmark, Norway, Sweden), CC-IN2P3 (France), KIT/GridKA (Germany), INFN-CNAF (Italy), NL-T1 (Netherlands), PIC (Spain), ASGC (Taiwan), RAL (UK) and BNL (USA), the Tier-2 facilities worldwide and large non-WLCG resource providers. Major contributors of computing resources are listed in Ref.~\cite{ATL-GEN-PUB-2016-002}.

%% file: atlas_authlist.tex
\begin{flushleft}
{\Large The ATLAS Collaboration}

\bigskip

M.~Aaboud$^\textrm{\scriptsize 136d}$,
G.~Aad$^\textrm{\scriptsize 87}$,
B.~Abbott$^\textrm{\scriptsize 114}$,
J.~Abdallah$^\textrm{\scriptsize 8}$,
O.~Abdinov$^\textrm{\scriptsize 12}$,
B.~Abeloos$^\textrm{\scriptsize 118}$,
R.~Aben$^\textrm{\scriptsize 108}$,
O.S.~AbouZeid$^\textrm{\scriptsize 138}$,
N.L.~Abraham$^\textrm{\scriptsize 152}$,
H.~Abramowicz$^\textrm{\scriptsize 156}$
% ...
\bigskip
\\
% ...
$^{8}$ Department of Physics, The University of Texas at Arlington, Arlington TX, United States of America\\
% ...
$^{12}$ Institute of Physics, Azerbaijan Academy of Sciences, Baku, Azerbaijan\\
% ...
$^{87}$ CPPM, Aix-Marseille Universit{\'e} and CNRS/IN2P3, Marseille, France\\
% ...
$^{108}$ Nikhef National Institute for Subatomic Physics and University of Amsterdam, Amsterdam, Netherlands\\
% ...
$^{114}$ Homer L. Dodge Department of Physics and Astronomy, University of Oklahoma, Norman OK, United States of America\\
% ...
$^{118}$ LAL, Univ. Paris-Sud, CNRS/IN2P3, Universit{\'e} Paris-Saclay, Orsay, France\\
% ...
$^{136}$ $^{(a)}$ Facult{\'e} des Sciences Ain Chock, R{\'e}seau Universitaire de Physique des Hautes Energies - Universit{\'e} Hassan II, Casablanca; $^{(b)}$ Centre National de l'Energie des Sciences Techniques Nucleaires, Rabat; $^{(c)}$ Facult{\'e} des Sciences Semlalia, Universit{\'e} Cadi Ayyad, LPHEA-Marrakech; $^{(d)}$ Facult{\'e} des Sciences, Universit{\'e} Mohamed Premier and LPTPM, Oujda; $^{(e)}$ Facult{\'e} des sciences, Universit{\'e} Mohammed V, Rabat, Morocco\\
% ...
$^{138}$ Santa Cruz Institute for Particle Physics, University of California Santa Cruz, Santa Cruz CA, United States of America\\
% ...
$^{152}$ Department of Physics and Astronomy, University of Sussex, Brighton, United Kingdom\\
% ...
$^{156}$ Raymond and Beverly Sackler School of Physics and Astronomy, Tel Aviv University, Tel Aviv, Israel\\
% ...
$^{a}$ Also at Department of Physics, King's College London, London, United Kingdom\\
$^{b}$ Also at Institute of Physics, Azerbaijan Academy of Sciences, Baku, Azerbaijan\\
$^{c}$ Also at Novosibirsk State University, Novosibirsk, Russia\\
$^{d}$ Also at TRIUMF, Vancouver BC, Canada\\
$^{e}$ Also at Department of Physics {\&} Astronomy, University of Louisville, Louisville, KY, United States of America\\
% ...
\end{flushleft}